\documentclass[journal]{IEEEtran}
\ifCLASSINFOpdf
\else
\fi

\usepackage{amsmath}
\usepackage{physics}
\usepackage{booktabs}
\usepackage{tikz}
\usepackage{cite}
\usepackage{multirow}
\usepackage{siunitx}
\usepackage{subfigure}
\usepackage{amsmath}
\usepackage{subcaption}
\usepackage{subcaption}
\usepackage{graphicx}

\usepackage{algorithm}
\usepackage{algorithmic}

\usetikzlibrary{matrix,calc}

\usepackage[varbb]{newpxmath}
\usepackage{xcolor}
\usepackage{changes}
\allowdisplaybreaks
\newcommand{\icms}{\textit{i}CMS}
\newtheorem{remark}{Remark}[subsection]
\DeclareMathOperator{\diag}{diag}
\DeclareMathOperator{\sgn}{sgn}
\newcommand{\probP}{\text{I\kern-0.15em P}}
\allowdisplaybreaks

\begin{document}
\setlength{\textfloatsep}{2pt}
%
% paper title
% Titles are generally capitalized except for words such as a, an, and, as,
% at, but, by, for, in, nor, of, on, or, the, to and up, which are usually
% not capitalized unless they are the first or last word of the title.
% Linebreaks \\ can be used within to get better formatting as desired.
% Do not put math or special symbols in the title.
\title{A Hierarchical Stochastic Model Predictive Control Framework for Integrated Request-aware Charge Scheduling and Service Allocation}
%
%
% author names and IEEE memberships
% note positions of commas and nonbreaking spaces ( ~ ) LaTeX will not break
% a structure at a ~ so this keeps an author's name from being broken across
% two lines.
% use \thanks{} to gain access to the first footnote area
% a separate \thanks must be used for each paragraph as LaTeX2e's \thanks
% was not built to handle multiple paragraphs
%

\author{Mainak~Dan,  Arvind~Easwaran,~\IEEEmembership{IEEE Senior Member}
\thanks{Mainak Dan is with Univ. Grenoble Alpes, CNRS, Grenoble INP, G2Elab, 38000 Grenoble, France, e-mail: mainak.dan@grenoble-inp.fr.}
\thanks{Arvind Easwaran is with the College of Computing and Data Science, Nanyang Technological University, 50 Nanyang Avenue, 639798, Singapore,  e-mail: arvinde@ntu.edu.sg.} 
%\thanks{This study is supported under the RIE2020 Industry Alignment Fund – Industry Collaboration Projects (IAF-ICP) Funding Initiative, as well as cash and in-kind contributions from the industry partner(s).}
}
\maketitle

% As a general rule, do not put math, special symbols or citations
% in the abstract or keywords.
\begin{abstract}
%% Text of abstract
\textcolor{black}{This study introduces a unified control mechanism for integrated charging and service allocation of Mobility-on-Demand Electric Vehicles (MoD-EVs) operating under a flexible A-to-B rental model.} The charging operation of MoD-EVs is influenced by stochastic customer rental requests and time-varying electricity prices. The framework captures the nonlinear battery dynamics and optimizes the charging operations against dynamic electricity prices and \textcolor{black}{to meet} stochastic customer demand while prioritizing battery health and cost efficiency. To \textcolor{black}{address the computational challenges posed by the} nonlinear battery dynamics, the framework employs a piece-wise linear model integrated into a multi-objective, chance-constrained mixed-integer linear programming (MILP) Model Predictive Control (MPC) formulation. A mixed logical switching mechanism is utilized to determine optimal charging sequences. Furthermore, a distributed approach is implemented to ensure computational scalability compared to centralized alternatives. Evaluation of this approach using a state-of-the-art commercial solver with stochastic EV rental requests under different confidence levels and time-varying electricity prices demonstrates significant benefits of the integrated mechanism design, including a reduction in charging costs and battery capacity degradation compared to the prevailing business-as-usual (BAU) approach and state of the art Laxity-based charging (LC) approach.

\end{abstract}

\IEEEpeerreviewmaketitle
\begin{color}{black}
\section*{Nomenclature}

\addcontentsline{toc}{section}{Nomenclature}
\subsection*{Abbreviations}
\begin{IEEEdescription}[\IEEEusemathlabelsep\IEEEsetlabelwidth{\mbox{$|\mathcal{RQ}^{\rm{cs}}[k]|$}}]
\item[MoD-EV] Mobility-on-Demand Electric Vehicle.
\item[ECM] Equivalent circuit model.
\item[SOC] State of charge.
\item[SOE] State of energy.
\item[OCV] Open circuit voltage.
\item[MILP] Mixed Integer Linear Programming.
\item[MPC] Model Predictive Control.
\item[\textit{i}CMS] Integrated charging management and service allocation mechanism. 
\end{IEEEdescription}

\subsection*{Set and Indices}
\begin{IEEEdescription}[\IEEEusemathlabelsep\IEEEsetlabelwidth{\mbox{$|\mathcal{RQ}^{\rm{cs}}[k]|$}}]
\item[$\mathcal{H}$] Set of time-bins. 
\item[$\mathcal{EC}$] The fleet of MoD-EVs.
\item[${\mathrm{ec}}$] Index of a MoD-EV in $\mathcal{EC}$.
\item[$\mu$] Set of charging modes $\{ {\rm{CM_1}}, {\rm{CM_2}}, {\rm{OFF}}, {\rm{DM}}\}$.
\item[$\mathcal{ST}$] Set of charging mode transitions. 
\item[$k$] Time index.
\end{IEEEdescription}

\subsection*{Decision and State Variables}
\begin{IEEEdescription}[\IEEEusemathlabelsep\IEEEsetlabelwidth{\mbox{$|\mathcal{RQ}^{\rm{cs}}[k]|$}}]
\item[$\mathbf{x}^{\rm{ec}}$] State vector of EV battery state-space model.
\item[$\mathbf{y}^{\rm{ec}}$] Output vector for EV battery state-space model.
\item[$i^{\rm{ec}}$] Current for charging/discharging operations.
\item[$\xi_c^{\rm{ec}}$] State of charge.
\item[$\xi_{avg}^{\rm{ec}}$] Average state of charge.
\item[$\xi_{dev}^{\rm{ec}}$] Deviation in state of charge.
\item[$T_{\rm{in}}^{\rm{ec}}$] Internal temperature of battery [\si{\kelvin}].
\item[$i^{\rm{ec}}_{R_j^{\rm{ec}}}$] Current through resistor $j$ of ECM [\si{\ampere}].
\item[$v^{\rm{ec}}$] Terminal voltage of battery [\si{\volt}].
\item[$e^{\rm{ec}}$] State of energy.
\item[$\delta^{{\rm{ec}},r_i}_{\rm{all}}$] Service allocation variable.
\item[$\delta_{c,l}^{\rm{ec}}$] Operational segment indicator in piecewise model.
\item[$\delta^{\rm{ec}}_{\mu}$] Charging mode indicator $\mu$.
\item[$n_{\rm{all}}^{\rm{cs}}$] Number of allocated EVs.
\item[$C_{\rm{el}}^{\rm{ec}}$] Electricity cost .
\item[$C_{\rm{deg}}^{\rm{ec}}$] Degradation cost
\item[$C_{\rm{sw}}^{\rm{ec}}$] Charging mode switching cost.
\item[$C_{\rm{cy}}^{\rm{ec}}$] Cyclic capacity loss of battery.

\end{IEEEdescription}

\subsection*{Parameters}
\begin{IEEEdescription}[\IEEEusemathlabelsep\IEEEsetlabelwidth{\mbox{$|\mathcal{RQ}^{\rm{cs}}[k]|$}}]
\item[$\Delta t$] Sample interval.
\item[$R_j^{\rm{ec}}$] ECM electrical resistors.
\item[$R_{\rm{th}}^{\rm{ec}}$] Thermal resistance of battery.
\item[$C_j^{\rm{ec}}$] ECM capacitance.
\item[$Q^{\rm{ec}}$] Capacity of the battery.
\item[$v_{\rm{OC}}^{\rm{ec}}$] Open circuit voltage.
\item[$\eta^{\rm{ec}}$] Charging and/ discharging efficiency.
\item[$\eta_{\rm{th}}^{\rm{ec}}$] Thermal efficiency.
\item[$k_1^{\rm{ec}}-k_4^{\rm{ec}}$] Cyclic capacity fading model parameters.
\item[$\textbf{A}_l^{\rm{ec}}, \textbf{B}_l^{\rm{ec}}$] State-space matrices of segment $l$.
\item[$R_{j,l}^{\rm{ec}}, C_{j,l}^{\rm{ec}}$] Modeling parameters of segment $l$.
\item[$v_{{\rm{OC}},l}^{\rm{ec}}$] OCV in segment $l$.
\item[$k_1^{\rm{ec}}$--$k_4^{\rm{ec}}$] Capacity degradation model parameters.
\item[$E_{\rm{dep}}^{\rm{ec}}$] Departure SOE.
\item[\mbox{$\mathcal{RQ}^{\rm{cs}}[k]$}] Set of requests at a charging station ${\rm cs}$ at time $k$.
\item[\mbox{$n_{\rm{rq}}^{\rm{cs}}$}] Number of ride requests.
\item[\mbox{$\alpha_{\rm{CON}}$}] Confidence level for chance constraint.
\end{IEEEdescription}

\end{color}

\section{Introduction}
\subsection{Motivation } \label{sec:motivationrelatedworks}
% Electrification of the transportation sector offers an intriguing alternative with many environmental and economic benefits in the context of global warming and the scarcity of fossil fuel \cite{knobloch2020net,reuters2020}. 
\IEEEPARstart{T}{he} increasing adoption of electric vehicles (EVs) for mobility-on-demand (MoD-EV) services such as ride-hailing and ride-sharing has led to a transition from personally owned vehicles and individual transportation towards rental-cum-sharing services provided by both private and public transportation companies in recent years \cite{zhao2023feasibility,xu2024demand,li2022evaluating}. To meet the ever-changing spatial and temporal patterns of customer demand for such services, particularly in urban areas, mobility providers will encounter difficulties in efficiently overseeing and arranging their expanding fleets of electric vehicles \cite{zeng2024economic}. The integration of EVs with the current energy infrastructure poses several inevitable difficulties, such as power outages, and excessive charging costs, especially if multiple EVs are charging without any coordinated strategies \cite{mahmud2023global,yu2022data,pankaj2022electric}. In addition, the heterogeneous characteristics of EVs and different energy requirements due to the battery size, state of health, and driving range (energy) requirements can create challenges for interoperability and coordination, leading to the inefficient use of charging infrastructure. Moreover, the inaccurate modeling of the batteries leads to insufficient energy storage, contributing to range anxiety, one of the intrinsic problems associated with EV users. In short, to address these \textcolor{black}{socio-economic} and technological challenges, several strategies encompassing the development of smart and advanced scheduling algorithms for automated and optimal charge scheduling \textcolor{black}{and service allocation} have become an utmost requirement.

\subsection{Related Works}

Research on EVs spans from strategies to minimize \textcolor{black}{charging} costs for individual owners to the comprehensive management of \textcolor{black}{operational costs} for large fleets, e.g. optimizing charging schedules, integrating renewable energy, developing vehicle-to-grid (V2G) technology, and exploring economic and policy implications to enhance the sustainability, reliability, and cost-effectiveness of electric transportation \cite{gnanavendan2024challenges,sadeghian2022comprehensive}. The existing literature that target modeling approaches can be categorized into two groups concerning EVs' applicability: one focusing on their use as distributed energy sources within the current energy network, and the other on their application as mobility-on-demand vehicles in transportation services.

A predominant group of studies extensively examines the impacts of EV charging on current energy networks. These includes the investigation of energy management methods and effect of power quality due to high penetration of EVs in a grid-connected network \cite{li2021coordinating, menghwar2024market}. For example, Pan et al. proposed a coordination control strategy to improve the stability and efficiency of standalone DC microgrids integrated with PV and EV charging, addressing the challenges of power fluctuations and voltage instability caused by these variable energy sources \cite{pan2023energy}. The authors in \cite{wang2024multi} introduced a multi-objective optimization model that coordinates EV charging in distribution networks to balance the interests of distribution network operators and EV owners, using an innovative I-AUGMENCON method to minimize costs and power loss while enhancing voltage stability and operational efficiency. The authors in \cite{hull2024techno} considered fleet electrification pace, grid constraint frequency, and trickle charging, to maximize net present value and reduce carbon emissions through integrated photovoltaic-energy storage-charging station systems. These approaches aim to enhance grid management systems through ancillary services, e.g. the utilization of EVs to reduce renewable energy curtailment \cite{fan2022bi} and efficient integration of EV movement model with charging infrastructure planning \cite{luke2021joint}, particularly during periods when EVs are in stationary condition (see \cite{li2023electric}, and the references therein). In addition, the optimized usage of EVs as distributed energy resources for demand response and demand side management have also been studied and explored \cite{kuppannagari2019approximate}. Zeynali \textit{et al.} developed a multi-objective optimization model for scheduling plug-in hybrid EV fleet with wind farms as a price-maker player in the day-ahead wholesale market, aiming to minimize costs and emissions while considering uncertainties in vehicle fleets and wind power \cite{zeynali2021hybrid}.  A real-time charging scheme for electric vehicle charging stations has been proposed in \cite{zanvettor2024stochastic} to manage uncertain vehicle demand and optimize participation in incentive-based demand response programs. However, these studies primarily have focused on energy management and used EVs as distributed energy resources, with less emphasis on their impacts on both energy and transportation services. Moreover, most of the EV modeling techniques utilized an oversimplified state-of-the-charge (SOC) based linear energy exchange model that has the inherent limitations of capturing the battery responses under different charging conditions and input stimuli e.g., aggressive current profiles, which may cause severe battery degradation effects. Thus adaptation of detailed battery modeling in devising optimal charge scheduling in parity with the operational and safety constraints is needed  \cite{park2020reinforcement}.

Integrating charge scheduling with service allocation in rental-cum-sharing/mobility-on-demand services can significantly enhance both the quality of service and cost-efficiency for customers by leveraging the low electricity rates during off-peak times, anticipating high demand periods, and strategic service allocations. The research on mobility-on-demand pertains to comprehensive fleet operation and management, encompassing charge scheduling, order dispatching, and vehicle rebalancing, particularly for large-scale shared EV fleet operators \cite{yu2023coordinating, chen2023electric}. There exist research works in the direction of efficient routing problems to improve energy efficiency and to reduce driving range anxiety \cite{li2021mixed, zhou2022electric}. Chen \textit{et al.} have proposed a coordinated optimization framework for logistic EVs incorporating simultaneous loading and partial charging, renewable energy integration, and electricity market participation to enhance logistics efficiency, reduce costs, and minimize carbon emissions under uncertain electricity prices and renewable generation \cite{chen2024coordinated}. However, flexible charging options and applicability on heterogeneous fleets have not been explored.  In \cite{huang2023robust}, Huang \textit{et al.} proposed a robust optimization model for charging coordination of the electric buses under the uncertainty of remaining capacity, to minimize the total charging cost considering two modes of charging namely battery swapping, and plug-in charging. The authors in \cite{yi2021framework} have proposed a centralized integrated charging and dispatching strategy to simulate ride-hailing services in New York City to improve fleet performance. However, centralized approaches tend to show scalability issues in implementation for large-scale EV fleets. Tucker \textit{et al.} \cite{tucker2019online} have studied an online heuristic algorithm to maximize profit by optimizing the between-ride schedules of the EVs to exploit cheaper time-of-use electricity rates and behind-the-meter solar generation in addition to efficiently re-balancing the vehicles throughout the city to avoid congestion at charging facilities and pick-up locations. Liang \textit{et al.} \cite{liang2020mobility} designed a partially observable Markov decision process and utilized a blend of deep reinforcement learning and binary linear programming to craft a nearly optimal solution. In \cite{boewing2020vehicle}, the authors developed a MILP model followed by a heuristic algorithm framework for overall fleet operation and charge scheduling decisions. In \cite{turan2020dynamic}, the authors utilized dynamic programming and deep reinforcement learning to generate near-optimal control policies for the joint routing and charging problem under the stochastic nature of trip demands, renewable energy availability, and electricity prices.

\subsection{Research Gap} \label{sec:researchgap}

Building on the discussion of related works, it is evident that most research on MoD-EV charge scheduling and charging operation optimization relies on simplified SOC-based EV battery models and aggregated degradation cost-based capacity fading models. While these studies focus on broader aspects of fleet operations and management planning, they provide limited insights into battery behavior and the detailed mechanisms for selecting charging options under dynamic electricity pricing and uncertain customer demands. The inclusion of battery degradation in EV charging is essential, as existing demand-aware charge planning models for MoD-EVs overlook critical factors such as degradation-dependent parameters and the long-term impact of charging/discharging cycles on heterogeneous fleet performance \cite{Vermeer2022}. Additionally, existing literature has not addressed comprehensive charge planning that accounts for fulfilling rental requests. With the growing penetration of high-power ultra-fast and super-fast charging infrastructure, it has become crucial to consider the nonlinear dynamics such as current-terminal voltage response for safe operational optimization and capacity fading factors such as input current and internal temperature \cite{park2020reinforcement}. 

Following this argument, limited efforts have been dedicated in addressing the following issues in \textcolor{black}{the context of} MoD-EVs:

\begin{enumerate}
    \item \textcolor{black}{development of integrated charging mechanism considering service allocation under stochastic rental requests and nonlinear dynamic behaviour of EV batteries;}
    \item adaptation of dynamical characteristics-based capacity fading model into the EV battery charging coordination instead of an agglomerated cost model;
    \item consideration of stochastic and intermittent behavior of the customer rental requests and dynamic charging costs that impact the overall operational cost of a fleet operator.
\end{enumerate}

\subsection{Problem Statement and Contributions} \label{sec:contributions}
 The primary objective of this study is to address the limitations above for MoD-EVs operating on an A-to-B model in an urban setting, where the customer (who is also the driver) rents a self-driving EV from a charging station (CS) and returns to another. The EVs are parked and connected to a specific source CS, with demand reflecting customer-initiated rental requests. It is to be noted that the trip bookings although mentioning the destination CS, do not quantify or produce information about the probable distance travelled or route taken by the customer during the rented period. The core challenges of the fleet operator lies not only in matching a vehicle with an appropriate energy level to a rent request but also in planning the charging operation of all other vehicles to maximize service level quality for future rent requests which have the intrinsic nature of uncertainties. A fleet operator considers all the information related to the vehicular state of charge, electricity price, and future demands, and processes accordingly to devise charge scheduling to achieve the following objectives: 
\begin{enumerate}
    \item minimization of electricity cost for charging under dynamic electricity pricing structure;
    \item minimization of the batteries' capacity fading through strategic energy exchange with the grid;
    \item management of supply-demand imbalances by managing the uncertainties in customer-requests prediction through a chance-constrained receding horizon optimization framework.
\end{enumerate}

\begin{figure*}[h]
\centering
\includegraphics[width=\textwidth]{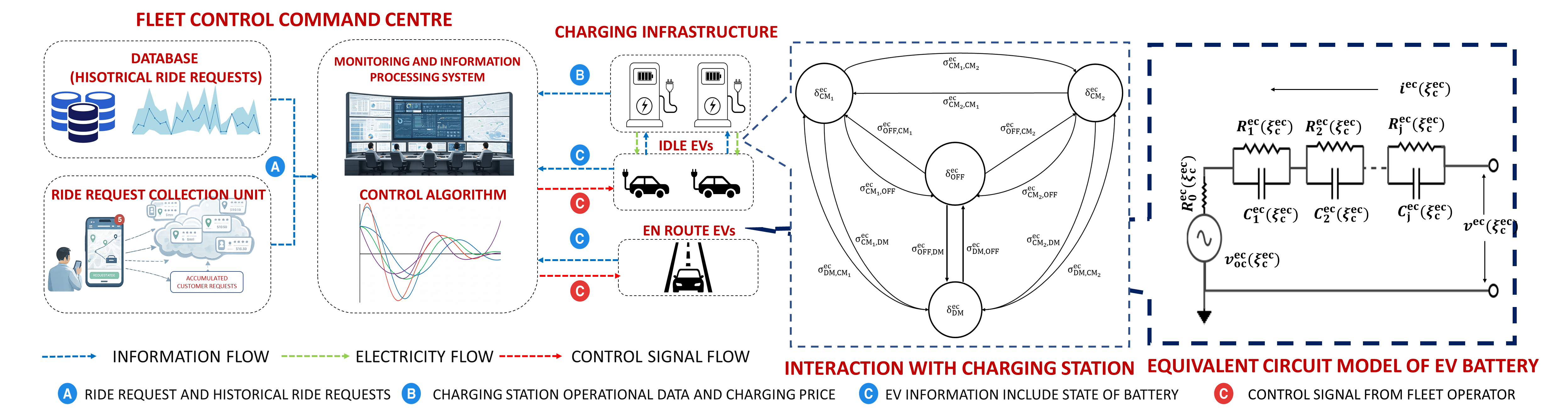}
\caption{Centralized Architecture and System Components}
\label{fig:charArch}
\end{figure*}

\textcolor{black}{
Ride-hailing systems exhibit uncertainty due to stochastic ride-request arrivals and variable travel conditions. Ride-request uncertainty is modeled using a scenario-based chance-constrained formulation constructed from historical ride-request data. By directly using empirical demand realizations, the proposed approach captures observed demand variability without assuming a specific parametric distribution. Service-level constraints are enforced probabilistically by requiring feasibility across a prescribed fraction of demand scenarios, corresponding to a selected confidence level. This confidence level controls the trade-off between operational reliability and system cost, with higher values yielding stronger service guarantees at the expense of increased resource utilization. This work focuses on optimizing service allocation and charging decisions under demand uncertainty rather than on demand forecasting. Accordingly, we assume that sufficiently accurate short-term ride-request forecasts are available, which is reasonable for the operational horizons considered.
}

The core accomplishment of this study is the development of an integrated charging management and service allocation mechanism (\icms) that addresses nonlinearities associated with EV batteries, customer-initiated rent requests, battery operational boundaries, and degradation. The key contribution of this research can be summarized as follows:

\begin{enumerate}
    \item adaptation of a comprehensive parametric battery model that captures and mirrors the nonlinear dynamical behavior of EV battery pack; 
    \item construction of a MILP problem through piece-wise linear relaxation of the nonlinear battery model for achieving improvement in computational efficiencies in the decision-making process;
    \item development of discrete charging operations and formulation of the transition automata among different charging modes in the presence of a unified charging model for vehicle-to-grid interaction;
    \item development of an allocation mechanism based upon the stored charge condition followed by a stochastic multi-objective model predictive control (MPC) framework for overall fleet charging and degradation cost minimization.
    \item development of a hierarchical distributed algorithm for overall integrated charge scheduling and ride allocation to reduce the computational complexity of the centralized approach.
\end{enumerate}

\subsection{Organization}

\textcolor{black}{The paper is organized as follows. Section \ref{sec:syscomponent} introduces the system components and dynamics forming the basis of the problem formulation. Section \ref{sec:MPCframework} formalizes the centralized stochastic optimization for fleet management. Section \ref{sec:distributed} presents a hierarchical distributed solution via two-stage stochastic optimization, splitting the problem into global planning and local scheduling. Section \ref{sec:casestudy} details the case study and experimental setup, and Section \ref{sec:conclusion} concludes with key findings and future research directions.}

\section{System Components and Dynamics} \label{sec:syscomponent}

\textcolor{black}{The primary objective of this investigation is to develop an integrated charging management and service allocation mechanism as shown in Fig. \ref{fig:charArch}, while simultaneously meeting operational and service level constraints. The centralized framework comprises of four key components: (i) battery dynamics and operational behavior, (ii) service allocation and ride-demand management, and (iii) vehicle-to-grid interaction.} Let us consider a fleet of $m$ EVs defined by the set, $\mathcal{EC} = \{{\rm{ec}}_1,{\rm{ec}}_2,\dots,{\rm{ec}}_m\}$. In addition, let us consider that the prediction window is divided into $H$ number of time-bins i.e. $\mathcal{H} = \{1, \dots, H \}$ with each having a duration $\Delta t$. The model captures either a vehicle's presence at a CS or the vehicle's expected arrival time through a vector $ \mathbf{Arr}^{\rm{ec}} \subset \{0,1\}^{ 1 \times H}$ where each entry defines the presence or absence of the vehicle ${\rm{ec}}$ at time-bin $k$ in the CS under consideration. It is to be noted that the set $\mathcal{EC}$ only defines the vehicle set that is present at the CS under consideration or en route to that CS and a subset of the actual fleet.

\subsection{Electric Vehicle Battery Dynamics Modeling}

Battery modeling is central to EV charge scheduling, as accurately capturing nonlinear battery behavior—due to varying discharge rates, temperature effects, and aging—enables reliable state estimation and efficient system operation \cite{park2020reinforcement}. EV batteries are commonly modeled using electrochemical, parametric, or data-driven approaches; however, electrochemical models are computationally intensive for large-scale optimization, and data-driven models require extensive data. \textcolor{black}{Alternatively, this study employs an enhanced equivalent circuit model (ECM) owing to its computational efficiency, scalability, and ability to accurately capture the trade-off between model accuracy and computational cost by effectively approximating both the transient and steady-state dynamics of batteries} \cite{6862541,hu2012comparative}.

\subsubsection{State Space Modeling} \label{subsec:evnonlinearmodeling}
In ECM approach, each battery is represented as an electric circuit that contains a voltage source (reflects the open circuit voltage (OCV)) $v_{\rm{OC}}^{\rm{ec}}$ in series with a resistor $R_0^{\rm{ec}}$, and a finite number of parallel resistor-capacitor (RC) $\{ (R_j^{\rm{ec}}, C_j^{\rm{ec}})\}, j = 1,\cdots,n$ components as depicted in Fig.~\ref{fig:charArch}. Let us consider the state vector of the vehicle battery as,
\begin{equation} \label{eq:statevariables}
    \mathbf{x}^{\rm{ec}} [k] = \begin{bmatrix}
                  T^{\rm{ec}}_{\rm{in}}[k],
                  \xi^{\rm{ec}}_c[k],
                  i^{\rm{ec}}_{R_1^{\rm{ec}}} [k],
                  i^{\rm{ec}}_{R_2^{\rm{ec}}} [k],
                  \dots\\
    \end{bmatrix}^\top,
\end{equation}
where, $T^{\rm{ec}}_{\rm{in}}[k]$ is the battery internal temperature, $\xi^{\rm{ec}}_c[k]$ is the state of the charge (SOC) defined as the measurement of available charge relative to the actual capacity, and $i^{\rm{ec}}_{R_j^{\rm{ec}}}[k], j = 1,\cdots,n$ is current through the resistor of the $j^{th}$ RC circuit at time instant $k$. \textcolor{black}{The output of the state space model is considered as the state of energy (SOE) $e^{\rm{ec}}[k]$ and the terminal voltage $v^{\rm{ec}}[k]$ at time instant $k$, which is expressed as a combination of OCV and, the voltage drops across the resistor components. It is to be noted that OCV and the model components are dependent on the SOC of the battery, i.e.,}
\begin{align} \label{eq:variableparameters}
    v_{\rm{OC}}^{\rm{ec}} = v_{OC}^{\rm{ec}} (\xi^{\rm{ec}}_c),~R_j^{\rm{ec}} = R_j^{\rm{ec}} (\xi^{\rm{ec}}_c), ~C_j^{\rm{ec}} = C_j^{\rm{ec}} (\xi^{\rm{ec}}_c), ~\forall j. 
\end{align}
Thus, the state estimation and output prediction equations become a nonlinear function of the state vector and input/output current $i^{\rm{ec}}[k]$ as follows.
\begin{align} 
    \mathbf{x}^{\rm{ec}}[k+1] &= f_x \left( \mathbf{x}^{\rm{ec}}[k] , i^{\rm{ec}}[k] \right), \label{eq:nonlineardynamics_1} \\
    \textcolor{black}{\mathbf{y}^{\rm{ec}}[k+1] = [e^{\rm{ec}}[k],v^{\rm{ec}}[k]]^\top} &\textcolor{black}{ ~= \mathbf{f_y} \left( \mathbf{x}^{\rm{ec}}[k] , i^{\rm{ec}}[k] \right). \label{eq:nonlineardynamics_2}}
\end{align}

\subsubsection{Capacity Fade Modeling}

\textcolor{black}{Capacity fading in batteries can be broadly classified into calendric aging and cyclic aging \cite{lam2012practical}. Calendric aging dominates in infrequently used vehicles, whereas cyclic aging is more pronounced in high-utilization applications such as rental fleets and MoD-EVs due to repeated charge–discharge cycles. Accordingly, we adopt the cyclic capacity fading model from \cite{lam2012practical}, given by,}
\begin{align} \label{eq:cycdeg}
    C_{\rm{cy}}^{\rm{ec}} =\left( k_1^{\rm{ec}} \xi_{dev} ^{\rm{ec}} e^{{k_2^{\rm{ec}} . \xi^{\rm{ec}}_{avg}}} + k_3^{\rm{ec}} e^{ {k^{\rm{ec}}_4 . \xi^{\rm{ec}}_{dev}}} \right). i^{\rm{ec}}\Delta t ,
\end{align}
where, $k_1^{\rm{ec}}-k_4^{\rm{ec}}$ are the modeling parameters. The stress factors during charging are the average SOC $\xi^{\rm{ec}}_{avg}$, SOC deviation $\xi^{\rm{ec}}_{dev}$, and amount of transferred charge $i^{\rm{ec}}$. The factors, $\xi^{\rm{ec}}_{avg}$ and $\xi^{\rm{ec}}_{dev}$ are computed as follows:
\begin{equation} \label{eq:socavg}
\begin{split}
    \xi_{avg}^{\rm{ec}}[k] &=\xi^{\rm{ec}}_c[k]  + 0.5 \times \eta^{\rm{ec}} \frac{i^{\rm{ec}}[k]\Delta t}{Q^{\rm{ec}}},\\
    \xi^{\rm{ec}}_{dev}[k]  &=0.5 \times \eta^{\rm{ec}} \frac{i^{\rm{ec}}[k] \Delta t}{Q^{\rm{ec}}}.
\end{split}
\end{equation}

\subsection{Service Allocation Mechanism}

The service allocation mechanism aims to match idle vehicles with incoming customer rent requests. Unlike personal EV charging, where the departure time is predetermined by the owner and known before scheduling, this investigation determines the EV's departure time using an automated allocation mechanism based on SOE. 
The set of incoming requests at the time-instant $k$ at the origin charging station is denoted as $\mathcal{RQ}^{\rm{cs}}[k] = \{ r_1, r_2,\cdots \}$.
Thus an EV $\rm{ec} \in \mathcal{EC}$ is allocated if the following expression holds,
\begin{subequations} \label{eq:ready1}
    \begin{align} 
      \delta^{{\rm{ec}},r_i}_{\rm{all}}[k] = 1 &\implies e^{\rm{ec}}[k] \ge {E}_{\rm{dep}}^{\rm{ec}} \cap \delta^{\rm{ec}}_{\rm{cs}}[k] = 1,\\
      & \textcolor{black}{\sum_{r_i} \delta^{{\rm{ec}},r_i}_{\rm{all}}[k] \le 1, \forall {\rm{ec}},}
\end{align}
\end{subequations}
where ${E}_{\rm{dep}}^{\rm{ec}}$ is a parameter denoting the predetermined departure energy level and the boolean variables $\delta^{{\rm{ec}},r_i}_{\rm{all}}[k], \delta^{\rm{ec}}_{\rm{cs}}[k] \in \{ 0,1\}$ denotes whether the vehicle is assigned at a time instant $k$ to the request $r_i$ and whether the car is present at the concerned charging station. 

Let us consider that $n_{\rm{all}}^{\rm{cs}}[k]$ is the number of vehicles that are allocated to the incoming service requests at time instant $k$. \textcolor{black}{In deterministic scenario, the following constraint needs to be satisfied to meet the customer ride-requests,}
\begin{equation} \label{eq:demandrequest2}
    n_{\rm{all}}^{\rm{cs}}[k] = \sum_{\rm{ec} \in \mathcal{C}} \delta^{\rm{ec}}_{\rm{all}}[k] \ge n_{\rm{rq}}^{\rm{cs}}[k] = |\mathcal{RQ}^{\rm{cs}}[k]|, ~ \forall k.
\end{equation}
\textcolor{black}{However, the number of ride requests at time instant $k$, denoted by $n_{\mathrm{rq}}^{\mathrm{cs}}[k]$, is inherently uncertain and cannot be predicted with perfect accuracy in advance. To account for this uncertainty, we adopt a scenario-based chance-constrained modeling framework derived from historical ride-request data. Instead of imposing a parametric assumption on the demand distribution, historical demands are directly used to characterize uncertainty, which is particularly suitable for ride-hailing systems exhibiting nonstationary, skewed, or multi-modal demand behavior. Consequently, service-level requirements are enforced through the following chance constraint:}
\begin{equation} \label{eq:chancconstraint}
     \mathbb{P} \left\{ n_{\rm{all}}^{\rm{cs}}[k] \ge n_{\rm{rq}}^{\rm{cs}}[k] \right\} \ge 1-\alpha_{\rm{CON}}, ~ \forall k.
\end{equation}
\textcolor{black}{where $\alpha_{\mathrm{CON}} \in [0,1]$ is the confidence level denoting the allowable risk of unmet demand and is treated as a design parameters that govern the trade-off between operational reliability and system cost.}

\subsection{Vehicle-to-Grid (V2G) Interaction Mechanism}
\textcolor{black}{The V2G interaction within a unified charging infrastructure is formulated by specifying the availability of multiple charging modes and modeling the mechanisms governing transitions between them. This unified framework is essential for ensuring flexibility and scalability, as it enables the seamless integration and coordinated management of diverse charging modes.} In this study, four discrete operational modes namely ${\rm{CM_1}}$ (charging), ${\rm{CM_2}}$ (charging), ${\rm{OFF}}$ and ${\rm{DM}}$ (discharging) are considered. The switching mechanism is general and can be extended to accommodate any number of charging modes offered by the charging infrastructure, enhancing the system's adaptability and efficiency in responding to diverse energy needs. These modes have been characterized by four mutually exclusive logical variables $\delta^{\rm{ec}}_{\mu} \in \{0,1\}$ with $\mu \in \{ {\rm{CM_1}}, {\rm{CM_2}}, {\rm{OFF}}, {\rm{DM}}\}$. Fig \ref{fig:charArch} depicts the transition from one mode to the other.

The four discrete modes $\delta^{\rm{ec}}_{\mu}[k]$ with $\mu \in \{ {\rm{CM_1}}, {\rm{CM_2}}, {\rm{OFF}}, {\rm{DM}}\}$ are characterized by the operating current at any time $k$. Specifically, when the EV is connected to the charging grid, at time instant $k$, it will either extract from or feed in energy to the grid. Depending upon the mode of exchange, the corresponding logical variable will be considered. In other words,
\begin{equation} \label{eq:operatingcurrent}
    \begin{cases}
      i^{\rm{ec}}[k] = 0 & \iff \delta^{\rm{ec}}_{\rm{OFF}}[k] = 1,\\
       \underline{i}^{\rm{ec}}_1 \le i^{\rm{ec}}[k] \le  \bar{i}^{\rm{ec}}_1 & \iff \delta^{\rm{ec}}_{\rm{CM_1}}[k] = 1,\\
      \underline{i}^{\rm{ec}}_2 \le i^{\rm{ec}}[k] \le  \bar{i}^{\rm{ec}}_2 & \iff \delta^{\rm{ec}}_{\rm{CM_2}}[k] = 1,\\
      -\underline{i}^{\rm{ec}}_3 \le i^{\rm{ec}}[k] \le  -\bar{i}^{\rm{ec}}_3 & \iff \delta^{\rm{ec}}_{\rm{DM}}[k] ~= 1,
    \end{cases}       
\end{equation}

\begin{equation} \label{eq:single_operation_mode}
    \delta^{\rm{ec}}_{\rm{OFF}}[k] + \delta^{\rm{ec}}_{\rm{CM_1}}[k]+ \delta^{\rm{ec}}_{\rm{CM_2}}[k]+\delta^{\rm{ec}}_{\rm{DM}}[k] =1.
\end{equation}
The transition among different operational modes as shown in Fig. \ref{fig:charArch} can be expressed with the help of the following expressions,
\begin{equation} \label{eq:operatingcurrent3}
    \sigma_{\mu_1,\mu_2}^{\rm{ec}}[k] = \delta_{\mu_1}^{\rm{ec}}[k-1] \wedge \delta_{\mu_2}^{\rm{ec}}[k],
\end{equation}
where $\sigma_{\mu_1,\mu_2}^{\rm{ec}} \in \{0,1\}$ with $(\mu_1,\mu_2) \in \mathcal{ST}$ with
\begin{equation} \label{eq:setstatetransitions}
    \mathcal{ST} \in \left\{ \begin{array}{cc}
        ({\rm{OFF,CM_1}}), ({\rm{CM_1,OFF}}), ({\rm{OFF,CM_2}}), \\
        ({\rm{CM_2,OFF}}), ({\rm{CM_1,CM_2}}), ({\rm{CM_2,CM_1}}), \\
        ({\rm{OFF,DM}}), ({\rm{DM,OFF}}), ({\rm{DM,CM_1}}), \\
        ({\rm{CM_1,DM}}), ({\rm{DM,CM_2}}), ({\rm{CM_2, DM}})
    \end{array} \right \}
\end{equation}
\section{Optimization Problem Formulation} \label{sec:MPCframework}

This section aims to establish an optimal control framework for \icms~that \textcolor{black}{addresses and devises a scalable approach to incorporate the nonlinear behavior of EV batteries as discussed in eqs. \eqref{eq:nonlineardynamics_1} and \eqref{eq:nonlineardynamics_2} and the logical constraints of V2G interactions as discussed in eqs. \eqref{eq:operatingcurrent}-\eqref{eq:operatingcurrent3}}. Incorporating nonlinear dynamics into the framework presents challenges for real-time applications, particularly for large EV fleets.  \textcolor{black}{Thus, the central challenge lies in translating the dynamic and logical behavior of system components into an MILP formulation and in developing a chance-constrained model predictive control (CCMPC) framework to enable request-aware charging planning for a fleet of MoD-EVs under time-varying electricity prices.}

\subsection{EV Battery Dynamics Approximation}
% Following the discussion on battery dynamics in section \ref{subsec:evnonlinearmodeling}, the modeling parameters display nonlinear behavior across the entire operational state of charge (SOC) range. 
The EV batteries show non-linear responses with regards to their SOC across the operational range due to variation in modeling parameters (ref. to eq. \eqref{eq:variableparameters}) as experimentally demonstrated and depicted later in section \ref{sec:sysdescription} and Fig. \ref{fig:batteryparameters}(A). First, we introduce the linear dynamics model, assuming the parameters remain constant throughout the dynamic range. Subsequently, we adopt the concept of simplified linear dynamics to construct the piecewise approximation.  

\subsubsection{Battery Linear Dynamics} \label{sec:lineardynamics}

\color{black}
Following the definition of state variables in eq. \eqref{eq:statevariables}, the vectorized representation of the linear battery state-space dynamics can be written as \cite{6862541,hu2012comparative},
\begin{equation} \label{eq:batteryspacestates}
    \mathbf{x}^{\rm{ec}}[k+1] = \textbf{A}^{\rm{ec}} \mathbf{x}^{\rm{ec}}[k] + \textbf{B}^{\rm{ec}} i^{\rm{ec}}[k], 
\end{equation}
where, $\textbf{A}^{\rm{ec}}$ and $\textbf{B}^{\rm{ec}}$ are state matrices. By definition, 
\begin{align} \label{eq:matrices}
&\textbf{A}^{\rm{ec}} = \diag \left[ \eta_{\rm{th}}^{\rm{ec}},1,e^{ {-\frac{\Delta t}{\tau^{\rm{ec}}_1}}},e^{ {-\frac{\Delta t}{\tau^{\rm{ec}}_2}}},\cdots \right],&  \\  
    &\textbf{B}^{\rm{ec}} = \begin{bmatrix}
    R_{\rm{th}}^{\rm{ec}}*\sgn(i^{\rm{ec}}[k]),
    \frac{\eta^{\rm{ec}}  \Delta t}{Q^{\rm{ec}}},
    1-e^{ {-\frac{\Delta t}{\tau^{\rm{ec}}_1}}},
    1-e^{ {-\frac{\Delta t}{\tau^{\rm{ec}}_2}}},
    \cdots
    \end{bmatrix}^\top,& \nonumber
\end{align}
where $R_{\rm{th}}^{\rm{ec}}$ is the thermal resistance, $0 < \eta_{\rm{th}}^{\rm{ec}} < 1$ is a factor that captures the temperature decay effect if idle,  $\sgn(i^{\rm{ec}}) = +1$ if $i^{\rm{ec}} \ge 0$ and $\tau^{\rm{ec}}_j = R^{\rm{ec}}_j C^{\rm{ec}}_j$. 
The terminal voltage $v^{\rm{ec}}$, regarded as the output of the state-space model, considers OCV $v_{\rm{OC}}^{\rm{ec}}$, the voltage across the series resistance $R_0^{\rm{ec}} (\xi^{\rm{ec}}_c)$, and the resistors of the RC parallel circuits, 
\begin{equation} \label{eq:output1}
    %v^{\rm{ec}}[k] = v_{OC}^{\rm{ec}} (\xi^{\rm{ec}}_c) [k] + v^{\rm{ec}}_h[k] - \sum_j R^{\rm{ec}}_j i^{\rm{ec}}_{R^{\rm{ec}}_j} [k] - R^{\rm{ec}}_0 i^{\rm{ec}}[k]
    v^{\rm{ec}}[k] = v_{\rm{OC}}^{\rm{ec}}[k] + R^{\rm{ec}}_0 i^{\rm{ec}}[k] + \sum_j R^{\rm{ec}}_j i^{\rm{ec}}_{R^{\rm{ec}}_j} [k].
\end{equation}
The SOE can be modeled as a linear function of SOC as follows,
\begin{equation} \label{eq:output2}
    e^{\rm{ec}}[k] = \theta^{\rm{ec}} \xi^{\rm{ec}}_c[k].
\end{equation}
\color{black}

\subsubsection{Piecewise Approximations}

Based on the linear dynamics described in section \ref{sec:lineardynamics}, this investigation reduces the computational complexity by dividing the operational SOC range of an EV battery into $\rm{L}$ mutually exclusive yet consecutive sections, each of which exhibits linear output characteristics in the OCV-SOC dynamics. Let $\delta^{\rm{ec}}_{c,l} \in \{0,1\}, l = 1,\dots,\rm{L}$, indicates the active operating segment such that the following condition is satisfied,
   \begin{align} \label{eq:pwpartition}
   \delta^{\rm{ec}}_{c,l} [k] \underline{\xi}_{c,l}^{\rm{ec}} \le \xi_c^{\rm{ec}}[k] < \delta^{\rm{ec}}_{c,l} [k] \bar{\xi}_{c,l}^{\rm{ec}},~~~ \sum_{l=1}^{\rm{L}} \delta^{\rm{ec}}_{c,l} [k] = 1. 
\end{align} 
In addition, to incorporate continuity between two consecutive segments, the following boundary condition is imposed,
\begin{equation} \label{eq:pwcontinuity}
    \underline{\xi}_{c,l}^{\rm{ec}} =  \bar{\xi}_{c,l-1}^{\rm{ec}}.
\end{equation}
The following condition which shows that the modeling parameters are fixed in values in each segment is imposed to obtain the piece-wise linear model,

\begin{equation} \label{eq:linearpwsections}
   \delta^{\rm{ec}}_{c,l} [k] = 1 \implies
    \begin{cases}
   v_{\rm{OC}}^{\rm{ec}}[k] = v_{{\rm{OC}},l}^{\rm{ec}},\\
      R_j^{\rm{ec}} = R_{j,l}^{\rm{ec}}, ~\forall j\\ C_j^{\rm{ec}} = C_{j,l}^{\rm{ec}}, ~\forall j.
    \end{cases}
\end{equation}
Utilizing eq. \eqref{eq:linearpwsections} in eq. \eqref{eq:matrices} and \eqref{eq:output1}, the state dynamics and output equation can be formulated through the following set of equations,
\begin{equation} \label{eq:statespace}
   \delta^{\rm{ec}}_{c,l} [k] = 1 \iff  \begin{cases}
     \mathbf{x}^{\rm{ec}}[k+1]  = 
      \textbf{A}_l^{\rm{ec}} \mathbf{x}^{\rm{ec}}[k] + \textbf{B}_l^{\rm{ec}} u^{\rm{ec}}[k], \\
     \small v^{\rm{ec}}[k] = 
       v_{{\rm{OC}},l}^{\rm{ec}}  + R^{\rm{ec}}_{0,l} i^{\rm{ec}}[k] + \sum_j R^{\rm{ec}}_{j,l} i^{\rm{ec}}_{R^{\rm{ec}}_{j,l}} [k],  \normalsize \\
     e^{\rm{ec}}[k] = 
       \theta^{\rm{ec}}_l \xi^{\rm{ec}}_{c,l}[k],
   \end{cases} 
\end{equation} 
where, $\theta^{\rm{ec}}_l$ is the transformation factor between the SOC and SOE.
\subsubsection{Battery Operational Constraints}

To cope with the expressions in \eqref{eq:statespace}, we consider two auxiliary variables $\phi^{\rm{ec}}_l(k)$ and $\chi^{\rm{ec}}_l(k)$ defined as,
\begin{equation}
\begin{cases}
  \phi^{\rm{ec}}_l(k) = \delta^{\rm{ec}}_{c,l} [k] \left( \textbf{A}_l^{\rm{ec}} \mathbf{x}^{\rm{ec}}[k] + \textbf{B}_l^{\rm{ec}} u^{\rm{ec}}[k] \right), \\
   \small \chi^{\rm{ec}}_l(k) = \delta^{\rm{ec}}_{c,l} [k] \left( v_{{\rm{OC}},l}^{\rm{ec}}  + R^{\rm{ec}}_{0,l} i^{\rm{ec}}[k] + \sum_j R^{\rm{ec}}_{j,l} i^{\rm{ec}}_{R^{\rm{ec}}_j,} [k] \right), \normalsize \\
    \mathbf{x}^{\rm{ec}}[k+1] = \sum_{l=1}^{\rm{L}}\phi^{\rm{ec}}_l(k),\\
    v^{\rm{ec}}[k] = \sum_{l=1}^{\rm{L}} \chi^{\rm{ec}}_l(k).
\end{cases}
\end{equation} 
\textcolor{black}{The following equations define the MIL constraints for the above expressions \cite{bemporad1999control}},

\small
\begin{equation}\label{eq:statespaces1}
\begin{cases}
  \phi^{\rm{ec}}_l(k) \le M_{\rm{PL}}\delta^{\rm{ec}}_{c,l} [k], \\
    \phi^{\rm{ec}}_l(k) \ge m_{\rm{PL}}\delta^{\rm{ec}}_{c,l} [k],\\
    \phi^{\rm{ec}}_l(k) \le  \textbf{A}_l^{\rm{ec}} \mathbf{x}^{\rm{ec}}[k] + \textbf{B}_l^{\rm{ec}} u^{\rm{ec}}[k] -m_{\rm{PL}}(1-\delta^{\rm{ec}}_{c,l} [k]), \\
    \phi^{\rm{ec}}_l(k) \ge  \textbf{A}_l^{\rm{ec}} \mathbf{x}^{\rm{ec}}[k] + \textbf{B}_l^{\rm{ec}} u^{\rm{ec}}[k] -M_{\rm{PL}}(1-\delta^{\rm{ec}}_{c,l} [k]);\\
    \chi^{\rm{ec}}_l(k) \le M_{\rm{PL}}\delta^{\rm{ec}}_{c,l} [k],\\
    \chi^{\rm{ec}}_l(k) \ge m_{\rm{PL}}\delta^{\rm{ec}}_{c,l} [k] ,\\
    \chi^{\rm{ec}}_l(k) \le v_{{\rm{OC}},l}^{\rm{ec}}  + R^{\rm{ec}}_{0,l} i^{\rm{ec}}[k] + \sum_j R^{\rm{ec}}_{j,l} i^{\rm{ec}}_{R^{\rm{ec}}_j} [k] -m_{\rm{PL}}(1-\delta^{\rm{ec}}_{c,l} [k]),\\
    \chi^{\rm{ec}}_l(k) \ge v_{{\rm{OC}},l}^{\rm{ec}}  + R^{\rm{ec}}_{0,l} i^{\rm{ec}}[k] + \sum_j R^{\rm{ec}}_{j,l} i^{\rm{ec}}_{R^{\rm{ec}}_j} [k] -M_{\rm{PL}}(1-\delta^{\rm{ec}}_{c,l} [k]); 
\end{cases}
\end{equation} \normalsize

\noindent where, $M_{\rm{PL}}$ and $m_{\rm{PL}}$ are the BigM operators. It is to be noted that eq. \eqref{eq:statespaces1}  approximates the nonlinear OCV-SOC response of EV batteries by introducing state matrices such as $\textbf{A}_l^{\rm{ec}}$ and $\textbf{B}_l^{\rm{ec}}$ that only model a particular linear section as defined in eq. \eqref{eq:pwpartition} and \eqref{eq:linearpwsections}.  
To find optimal results within the permissible and safe operating range of the EV batteries following constraints are also imposed,
\color{black}
\begin{equation} \label{eq:outputconstraints}
    \underline{\mathbf{x}}^{\rm{ec}} \le \mathbf{x}^{\rm{ec}}[k] \le \bar{\mathbf{x}}^{\rm{ec}},~~
   \underline{\mathbf{y}}^{\rm{ec}} \le \mathbf{y}^{\rm{ec}}[k] \le \bar{\mathbf{y}}^{\rm{ec}} .
\end{equation}
\color{black}

\subsection{Vehicle-to-Grid (V2G) Interaction Mechanism}
\textcolor{black}{In order to the integrate the V2G logical interaction and state transitions as described in eq \eqref{eq:operatingcurrent3} into the \icms~framework, the following MIL constraints are devised (following the discussion in \cite{bemporad1999control}),} 
\begin{align}\label{eq:operatingcurrent4}
    -\delta_{\mu_1}^{\rm{ec}}[k-1] + \sigma_{\mu_1,\mu_2}^{\rm{ec}}[k] &\le 0, \nonumber\\
    -\delta_{\mu_2}^{\rm{ec}}[k] + \sigma_{\mu_1,\mu_2}^{\rm{ec}}[k] &\le 0, \\
    \delta_{\mu_1}^{\rm{ec}}[k-1] + \delta_{\mu_2}^{\rm{ec}}[k] -  \sigma_{\mu_1,\mu_2}^{\rm{ec}}[k] &\le 1. \nonumber
\end{align} 
where $\sigma_{\mu_1,\mu_2}^{\rm{ec}} \in \{0,1\}$ with $(\mu_1,\mu_2) \in \mathcal{ST}$ with $\mathcal{ST}$ defined in eq. \eqref{eq:setstatetransitions}. While the aforementioned formulation designs the switching mechanism, we consider the total energy exchange between an EV and grid by defining a continuous variable $\Delta e^{\rm{ec}} [k]$ as,
\begin{align}
    \Delta e^{\rm{ec}}[k] &= e^{\rm{ec}}[k+1] - e^{\rm{ec}}[k], \label{eq:energyexchange}\\
    \delta_{\rm{DM}}^{\rm{ec}} [k] &= 1  \iff  \Delta e^{\rm{ec}} [k] < 0 \label{eq:energypositive}.
\end{align}
For ease of understanding, we will utilize the following variables in the rest of the investigation,
\begin{align*}
    \Delta e^{\rm{ec}} [k] = (\Delta e^{\rm{ec}}[k])^+, ~{\rm{if}}~\Delta e^{\rm{ec}} [k] \ge 0,\\
    \Delta e^{\rm{ec}} [k] = (\Delta e^{\rm{ec}}[k])^-, ~{\rm{if}}~\Delta e^{\rm{ec}} [k] < 0.
\end{align*}
To implement the CS energy limitations, we impose the following constraints,
\begin{align} \label{eq:csconstraints}
    (\Delta e^{\rm{ec}}[k])^+ \le \bar{E}^{\rm{cs}}, (\Delta e^{\rm{ec}}[k])^- \ge -\underline{E}^{\rm{cs}}.
\end{align}
\subsection{CCMPC Problem Formulation}
\begin{figure}[h]
\centering
\includegraphics[width=0.5\textwidth]{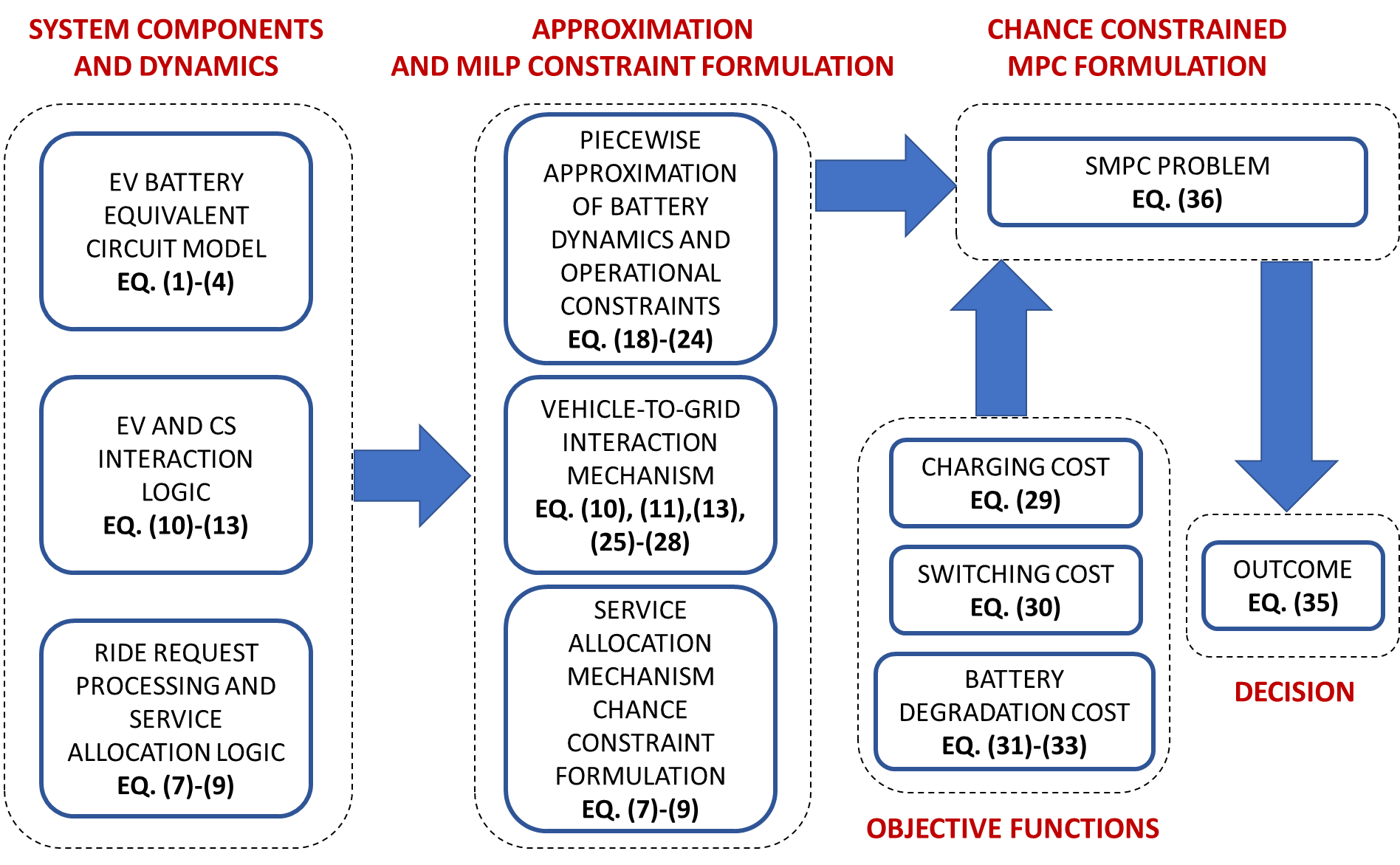}
\caption{\textcolor{black}{Centralized CCMPC Problem}}
\label{fig:centralized_smpc}
\end{figure}

The utilization of MPC is intended to incorporate time-dynamic electricity prices and customer requests, enabling request-aware charge planning for a fleet of vehicles that are either idle or en route to a specific CS. This section describes the associated costs and provides a detailed formulation of the MPC problem. \textcolor{black}{For ease of understanding, Fig.~\ref{fig:centralized_smpc} presents the approximation steps along with the associated constraint formulation and objective functions used to construct the CCMPC problem.}

\subsubsection{Operational Cost}

The total operational cost for the fleet charging includes - (i) the cost of electricity for charging, influenced by the dynamic electricity pricing, (ii) battery degradation cost, influenced by the energy exchange with the charging station (iii) switching cost, responsible for prohibiting abrupt switching of various charging modes and their inter-toggling. In this problem, we consider that the forecasted energy purchase and sale price is available for a particular planning horizon. The total electricity cost due to energy exchange with the grid can be formulated as,
\begin{equation} \label{eq:electricitycost}
    C_{\rm{el}}^{\rm{ec}}[k] = c_{\rm{el}}^{\rm{pur}}[k] (\Delta e^{\rm{ec}}[k])^+ - c_{\rm{el}}^{\rm{sale}}[k] (\Delta e^{\rm{ec}}[k])^- ,
\end{equation}
where $c_{\rm{el}}^{\rm{pur}}[k]$ and $c_{\rm{el}}^{\rm{sale}}[k]$ are the purchasing and feeding in unit energy price at time instant $k$.

Furthermore, considering the switching cost when choosing a charging mode is crucial, as it deters arbitrary shifts between charging strategies in the algorithm, a behavior that could be prevented by introducing additional expenses. The switching cost is defined as,
\begin{equation}
    C_{\rm{sw}}^{\rm{ec}}[k] = \sum_{(\mu_1,\mu_2) \in \mathcal{ST}} c^{\rm{ec}}_{\mu_1,\mu_2}  \sigma^{\rm{ec}}_{\mu_1,\mu_2} [k]
\end{equation}

The battery degradation cost can be calculated by projecting the unit capacity manufacturing cost upon the capacity loss during the charging-discharging cycle in order to implement the V2G system. The cyclic degradation as described in eq. \eqref{eq:cycdeg}-\eqref{eq:socavg} depends upon the amount of charge exchanged between the vehicle and the grid and the SOC condition. To tackle the complexity and high non-linearity of eq. \eqref{eq:cycdeg}-\eqref{eq:socavg}, we implemented mixed logical dynamic formulation \cite{bemporad1999control} to obtain the piece-wise linear approximation of the degradation function utilizing the binary variables $\delta^{\rm{ec}}_{c,l}[k]$ and $ \delta^{\rm{ec}}_{\mu}[k]$. The model constraints can be written as follows,
\begin{equation} \label{eq:cycdeglinear}
\begin{split}
      \gamma_{l,\mu}^{\rm{ec}}[k] = \delta^{\rm{ec}}_{c,l}[k] &\wedge \delta^{\rm{ec}}_{\mu}[k],  \\
\gamma_{l,\mu}^{\rm{ec}}[k] = 1 \iff  c_{\rm{deg}}^{\rm{ec}}[k] &= \mathcal{L} \left (  i^{\rm{ec}} [k], \xi^{\rm{ec}}_c [k] \right), 
\end{split}
\end{equation}
with
\begin{equation}
    \mathcal{L} \left (  i^{\rm{ec}} [k], \xi^{\rm{ec}}_c [k] \right) =  c_{0,l,\mu}^{\rm{ec}} + c_{1,l,\mu}^{\rm{ec}}  i^{\rm{ec}} [k] + c_{2,l,\mu}^{\rm{ec}} \xi^{\rm{ec}}_c [k],
\end{equation}
where, $\mu \in \{ {\rm{CM_1}}, {\rm{CM_2}}, {\rm{OFF}}, {\rm{DM}}\}$ and $l = 1,\dots,\rm{L}$. Considering $c^{\rm{man}}$ as the manufacturing cost of the battery per unit capacity, the overall degradation cost can be expressed as,
\begin{equation} \label{eq:finaldegradationcost}
    C_{\rm{deg}}^{\rm{ec}}[k] = c^{\rm{man}} c_{\rm{deg}}^{\rm{ec}}[k].
\end{equation}

\subsubsection{Problem Formulation}

The objective of \icms~is to minimize the multiple objectives occurring from the operational cost and the battery degradation cost. The comprehensive cost of preparing a vehicle for serving the customer request is calculated using a weighted sum of both operational cost as well as the battery degradation cost, as follows,
\begin{equation} \label{eq:objectivefunctionmultiple}
  C^{\rm{ec}} (\mathbf{w},\mathbf{d}^{\rm{ec}} ) [k] =  w_1 ( C_{\rm{el}}^{\rm{ec}}[k] +  C_{\rm{sw}}^{\rm{ec}}[k]) + w_2\lambda_{\rm{deg}} C_{\rm{deg}}^{\rm{ec}}[k],
\end{equation}
where, $\lambda_{\rm{deg}}$ acts as scaling factor and $w_1+w_2 = 1$. The vector of decision variables $\forall k,~\forall {\rm{ec}}$ is described as,
\begin{align} \label{eq:decisionvariable}
    \mathbf{d}^{\rm{ec}}[k] = & \left[ i^{\rm{ec}}[k],  \delta^{{\rm{ec}},r_i}_{\rm{all}}[k], \sigma_{\mu_1,\mu_2}^{\rm{ec}}[k],(\Delta e^{\rm{ec}}[k])^+, \right. \nonumber\\
    &\left. (\Delta e^{\rm{ec}}[k])^-, \gamma_{l,\mu}^{\rm{ec}}[k], \delta^{\rm{ec}}_{c,l}[k], \delta_{\mu}^{\rm{ec}}[k] \right]^\top,
\end{align}
with the vector of state variables being $\mathbf{x}^{\rm{ec}} [k]$ defined in \eqref{eq:statevariables}
\noindent and the output variable being $[v^{\rm{ec}}[k],e^{\rm{ec}}[k]]^\top$.
The chance-constrained MPC can be constructed by formulating the optimization problem for minimizing the weighted sum of the cost functions over a prediction horizon as follows,
%\
\begin{align} \label{eq:problemMPC}
     \mathcal{P_M}:~~&\min_{\mathbf{d}^{\rm{ec}}[k], \forall k, \forall {\rm{ec}}} \sum_{k \in \mathcal{H}} \sum_{{\rm{ec}} \in \mathcal{EC}} \mathbb{E}_{\mathcal{N}_{\rm{rq}}} \left\{ C^{\rm{ec}} ( \mathbf{w}, \mathbf{d}^{\rm{ec}} ) [k] \right\} \\
    {\rm{s.t.}},& ~ {\rm{constraints}}~\eqref{eq:chancconstraint}, \eqref{eq:pwpartition}-\eqref{eq:energypositive}, \forall k,\forall \rm{ec}. \nonumber
\end{align}

\textcolor{black}{The problem $\mathcal{P_M}$ in eq. \eqref{eq:problemMPC} defines the multi-time step, mixed-integer chance-constrained optimization problem with its structure illustrated in Fig. \ref{fig:centralized_smpc}. The methodology integrates four primary components: an objective function targeting electricity, switching, and degradation costs \eqref{eq:electricitycost}-\eqref{eq:finaldegradationcost}; mixed-integer piecewise-linear battery dynamics to ensure safe operation \eqref{eq:pwpartition}-\eqref{eq:outputconstraints}; V2G constraints governing bidirectional energy exchange and mode selection \eqref{eq:operatingcurrent}, \eqref{eq:single_operation_mode}, \eqref{eq:setstatetransitions} and \eqref{eq:operatingcurrent4}-\eqref{eq:csconstraints}; and a service allocation mechanism employing chance constraints \eqref{eq:ready1}-\eqref{eq:chancconstraint} to satisfy stochastic rental requests at defined confidence levels. The solution of the MPC problem yields a sequence of decisions $\mathbf{d}^{\rm ec}[k]$, defined in \eqref{eq:decisionvariable}, over the prediction horizon $\mathcal{H}$ for all vehicles ${\rm ec}\in\mathcal{EC}$. In line with the MPC paradigm, only the control actions at the current time step are implemented, after which the horizon is shifted and the problem is re-solved.}

\begin{remark}  \normalfont (Problem Complexity)
   Let's consider a fleet of $n_{veh}$ vehicles, which includes only the vehicles that are currently idle at the charging station or en route within a specified prediction horizon ($N$), while the number of operational states offered by the charging station is $n_s$. Each vehicle's battery dynamics are approximated using ${\rm{L}}$ piece-wise sections. Consequently, the number of binary decision variables required for the decision-making process at a specific time instant as modeled in section \ref{sec:MPCframework} is determined to be $ n_{var} = n_{veh}\left(3{\rm{L}}+1+n_s\left( 2{\rm{L}}+3+\frac{n_s-1}{2}\right)\right)$. Since problem \eqref{eq:problemMPC} entails a multi-time step decision-making process spanning $N$ time-steps, the total number of binary variables increases to $Nn_{var}$. Similarly, the number of continuous variables involved in the optimization problem can be calculated as $n_{veh} \left( N \left( n_s{\rm{L}}+10\right)+2 \right)$. Although there exist state-of-the-art MILP solvers such as branch-and-bound, the scale of the problem increases with the introduction of more EVs in the fleet to satisfy the increasing demand, and finding the solution to the centralized problem suffers from a slow convergence rate. Thus, to tackle the increasing scale of the problem, the solution method with improved scalability and lower computational requirements is needed for real-time implementation. 
\end{remark}

\section{Hierarchical \& Distributed Solution for \icms~} \label{sec:distributed}
In Section \ref{sec:MPCframework}, we demonstrated that the centralized problem's computational complexity increases with the fleet size growth and the prediction horizon extension. This increasing computational complexity poses significant challenges to real-time implementation due to scalability issues. This section presents a distributed hierarchical solution approach designed to effectively manage the computational complexity associated with the problem detailed in Eq. \eqref{eq:problemMPC}. The proposed method employs a two-stage hierarchical framework to solve the optimization problem efficiently: 

\begin{enumerate}
    \item \textbf{\textit{Upper (planning) stage:}} It employs a reduced order modeling, focusing on allocating the available vehicles to the customer requests by handling the chance constraints as described in eq. \eqref{eq:chancconstraint} over a horizon. The allocation mechanism results in determining the departure times of the idle EVs and thus provides an important parameter to find the solution in the next stage.

    \item \textbf{\textit{Lower (scheduling) stage:}} The lower stage employs decentralized optimization approach that decomposes the centralized problem into sub-problems for individual MoD-EVs, considering the constraints and objectives related to battery behavior, dynamic pricing, and departure time determined by the upper stage problem. The optimization at this level ensures that the EVs are charged adequately to meet their service commitments while minimizing overall operational costs.
\end{enumerate}

\subsection{Ride Assignment Problem}

The planning stage optimization problem deals with a high-level approach with a reduced number of decision variables. This stage aims to optimize the cost of charging to the desired level of SOE while maintaining the service level constraints over the prediction horizon. The problem can be defined as,
\begin{align} \label{eq:problemMPC2}
     \mathcal{P}_{{\rm{RA}}}:~~~~~~&\min_{\mathbf{d}_{\rm{PS}}^{\rm{ec}}[k], \forall k, \forall {\rm{ec}}}  \sum_{{\rm{ec}} \in \mathcal{EC}} \mathbb{E}_{\mathcal{N}_{\rm{rq}}} \left\{  \sum_{k \in \mathcal{H}}C_{\rm{el}}^{\rm{ec}}[k]  \right\}   \\
    {\rm{s.t.}},& ~ {\rm{constraints}}~\eqref{eq:ready1}, \eqref{eq:chancconstraint}, \eqref{eq:operatingcurrent}, \eqref{eq:output2}, \eqref{eq:energyexchange}, \eqref{eq:energypositive}, \eqref{eq:csconstraints}, \nonumber  \\
    &~{\rm{SOC:}}~\xi^{\rm{ec}}_c[k+1] = \xi^{\rm{ec}}_c[k] + \frac{\eta^{\rm{ec}}  \Delta t}{Q^{\rm{ec}}}i^{\rm{ec}}[k],\forall k,\forall \rm{ec}.   \nonumber
\end{align}

\textcolor{black}{The formulation incorporates chance-constrained service availability constraints \eqref{eq:ready1}–\eqref{eq:chancconstraint}, SOC dynamics and SOE \eqref{eq:output2} of MoD-EV, and energy exchange feasibility constraints \eqref{eq:energyexchange}–\eqref{eq:energypositive}, along with charging station capacity limits \eqref{eq:csconstraints}.}
The decision variables of this stage are defined as follows $ \forall k \in \mathcal{H}, \forall {\rm{ec}} \in \mathcal{EC}$, 
\begin{align} \label{eq:decisionvariable_ra}
    \mathbf{d}_{\rm{RA}}^{\rm{ec}}[k] = & \left[ \delta^{{\rm{ec}},r_i}_{\rm{all}}[k], (\Delta e^{\rm{ec}}[k])^+, (\Delta e^{\rm{ec}}[k])^-\right]^\top.
\end{align}
It is to be noted that the allocation variables $\delta^{\rm{ec}}_{\rm{all}}[k], \forall{\rm{ec}}, \forall{k}$ determine the departure time of EVs, $t^{\rm{ec}}_{\rm{dep}}, \forall{{\rm{ec}} \in \mathcal{EC}}$. This departure time is then transmitted to each idle EV for further scheduling of charging variables. \textcolor{black}{The details of the steps involved are discussed in Algorithm \ref{alg:RA}.}

\begin{algorithm}[h]
\color{black}
\caption{\textcolor{black}{Ride Assignment (RA) Optimization}}
\label{alg:RA}
\begin{algorithmic}[1]
\REQUIRE Prediction horizon $\mathcal{H}$; set of EVs $\mathcal{EC}$; historical demand profiles; electricity prices; station capacity limits; initial SOE of EVs; confidence level for chance constraints
\ENSURE Ride assignments $\mathbf{d}_{\rm{RA}}^{\rm{ec}}[k]$, and departure times $t^{\rm{ec}}_{\rm{dep}}$;
\STATE Define decision variables $\mathbf{d}_{\rm{RA}}^{\rm{ec}}[k]$ according to \eqref{eq:decisionvariable_ra}, $\forall k \in \mathcal{H}, \forall {\rm ec}\in\mathcal{EC}$;
\STATE Construct and solve problem $\mathcal{P}_{\rm{RA}}$ as defined in \eqref{eq:problemMPC2};
\RETURN $\mathbf{d}_{\rm{RA}}^{\rm{ec}}[k]$, $t^{\rm{ec}}_{\rm{dep}}$;
\STATE Communicate $t^{\rm{ec}}_{\rm{dep}}$ to idle EVs for subsequent charging-level scheduling.
\end{algorithmic}
\end{algorithm}

\begin{remark} \normalfont (Complexity)
The optimization problem in \eqref{eq:problemMPC2} reduces the number of discrete decision variables compared to \eqref{eq:problemMPC} focusing only on the high-level energy consumption and service allocation. 
\end{remark}
\begin{remark}\normalfont (Lower bound solution)
    The ride assignment problem serves as a lower bound solution to the charging cost of the original optimization problem in eq. \eqref{eq:problemMPC} by utilizing a simplified reduced order model while maintaining the essential energy consumption and station capacity constraints.
\end{remark}

\subsection{Charging Scheduling and Optimization(CO)}

The second stage optimization problem deals with the intricate EV battery dynamics and safe operational range, and the degradation costs by decomposition the problem into a multi-agent (EV) optimization problem. The application of multi-agent objective has three-fold advantages:
\begin{enumerate}
    \item Each optimization problem is solved by an individual EV (agent) bypassing the requirement of a centralized solver leading to a reduced computational burden.
    \item The departure time, $t^{\rm{ec}}_{\rm{dep}}, \forall{{\rm{ec}} \in \mathcal{EC}}$ within a pre-determined prediction horizon is only transmitted to the idle EVs standing at the CS leading to a fewer number of decomposed sub-problems to be solved in parallel contrary to the centralized approach which considers the en-route EVs reaching CS within that prediction horizon.
    \item Given a prediction horizon $H$, it is evident from the RA that $t^{\rm{ec}}_{\rm{dep}} \le H, \forall{{\rm{ec}} \in \mathcal{EC}}$, leading to a reduced number of optimization variables in each decomposed sub-problem.
\end{enumerate}

\textcolor{black}{Thus, in second stage,} the resultant optimization problem \textcolor{black}{for a MoD-EV} is a deterministic problem with a predetermined departure time as follows,
\begin{align} \label{eq:problemMPC3}
     \mathcal{P}^{\rm{ec}}_{{\rm{CO}}}:&\min_{\mathbf{d}_{\rm{SS}}^{\rm{ec}}[k], \forall k} \sum_{k =1}^{t^{\rm{ec}}_{\rm{dep}}}  C^{\rm{ec}} (\mathbf{w}, \mathbf{d}_{\rm{CO}}^{\rm{ec}} ) [k]+\lambda_{\rm{pen}} \left(   {E}_{\rm{all}}^{\rm{ec}}- e^{\rm{ec}}[t^{\rm{ec}}_{\rm{dep}}]  \right)^2 \nonumber \\
    {\rm{s.t.}},& ~ {\rm{constraints}}~\eqref{eq:pwpartition}-\eqref{eq:energypositive}, \forall k \in \mathcal{H},\forall \rm{ec} \in \mathcal{EC},   
\end{align}
with the decision variables for ${\rm{ec}} \in \mathcal{EC}, ~\forall{k} \in [1, t^{\rm{ec}}_{\rm{dep}} ]$ as follows,
\begin{align} \label{eq:decisionvariable_co}
    \mathbf{d}_{\rm{CO}}^{\rm{ec}}[k] = & \left[ i^{\rm{ec}}[k], \sigma_{\mu_1,\mu_2}^{\rm{ec}}[k],(\Delta e^{\rm{ec}}[k])^+, (\Delta e^{\rm{ec}}[k])^-, \right. \nonumber\\
    &\left. \gamma_{l,\mu}^{\rm{ec}}[k], \delta^{\rm{ec}}_{c,l}[k], \delta_{\mu}^{\rm{ec}}[k] \right]^\top,
\end{align}
and the vector of state variables being $\mathbf{x}^{\rm{ec}} [k]$ from \eqref{eq:statevariables}. \textcolor{black}{The details of the steps involved are discussed in Algorithm \ref{alg:CO}.}

\begin{algorithm}[]
\color{black}
\caption{\textcolor{black}{Decentralized Charging Optimization (CO)}}
\label{alg:CO}
\begin{algorithmic}[1]
\REQUIRE departure times $t^{\rm{ec}}_{\rm{dep}}$ from the RA stage for each MoD-EV in $\mathcal{EC}$; initial battery states $\mathbf{x}^{\rm{ec}}[0]$; electricity prices; battery and charging parameters
\ENSURE Optimal charging schedules $\mathbf{d}_{\rm{CO}}^{\rm{ec}}[k]$, $\forall {\rm ec}\in\mathcal{EC}$

\FOR{each idle EV ${\rm ec} \in \mathcal{EC}$ at the charging station}
    \STATE Set the prediction horizon to $[0, t^{\rm{ec}}_{\rm{dep}}-1]$, define the decision variables $\mathbf{d}_{\rm{CO}}^{\rm{ec}}[k]$ as in \eqref{eq:decisionvariable_co} and initialize the state vector $\mathbf{x}^{\rm{ec}}[0]$;
    \STATE Solve problem \eqref{eq:problemMPC3};
    \RETURN $\mathbf{d}_{\rm{CO}}^{\rm{ec}}[k]$, $\forall {\rm ec}\in\mathcal{EC}$;
\ENDFOR
\STATE Implement the charging action at the current time step

\end{algorithmic}
\end{algorithm}

\section{Case Study and Results} \label{sec:casestudy}

\subsection{Data Preprocessing} \label{sec:sysdescription}

\begin{figure}[h]
\centering
\includegraphics[scale = 0.44]{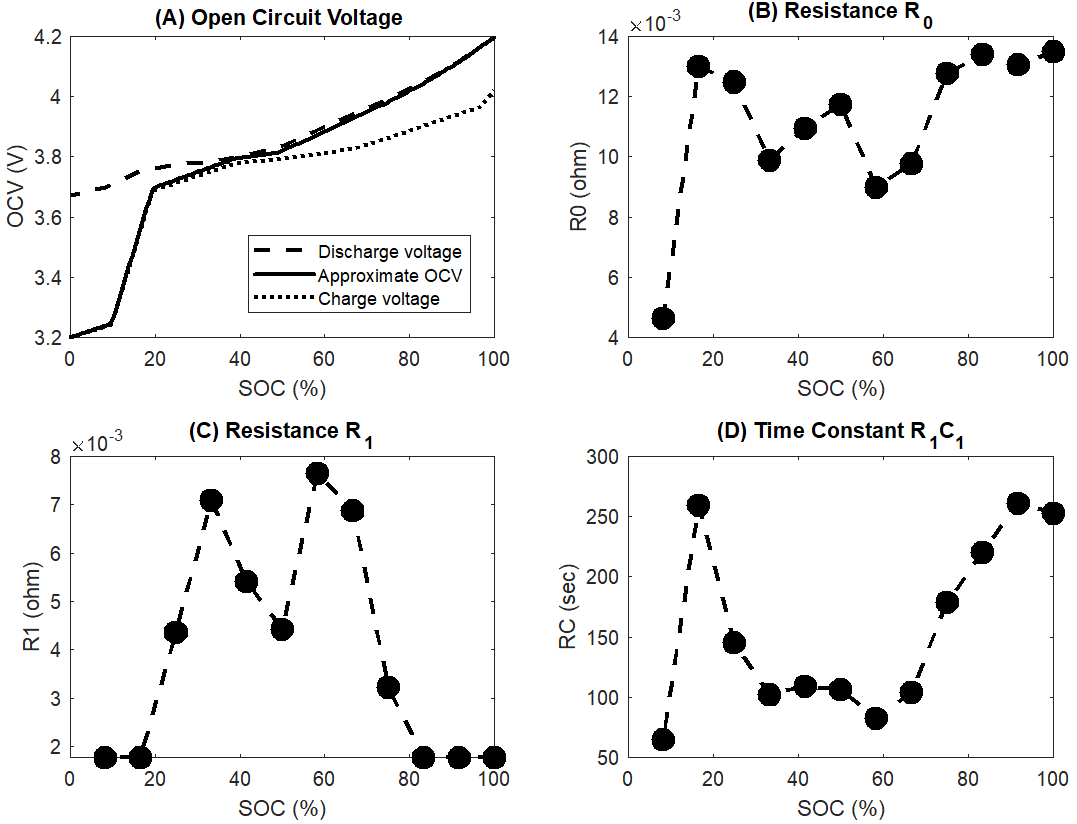}
\caption{Battery parameters and OCV modeling}
\label{fig:batteryparameters}
\end{figure}

\color{black}
To develop an accurate EV battery model, we utilized a dataset containing 18650 Li-ion cell information as described in \cite{bole2014adaptation}. In particular, the pulse discharge experimental dataset in \cite{bole2014adaptation} was used to identify the components of the equivalent circuit model depicted in Fig. \ref{fig:batteryparameters}, illustrating their behavior with SOC variations. Battery packs of different capacities were then constructed by combining these cells in series and parallel configurations, denoted as ($b_s, b_p$), to meet different energy requirements. While certain assumptions were made in scaling to pack-level models, the essential characteristics observed at the cell level were preserved, providing reliable representations of EV battery performance. The use of cell-level ageing parameters at the pack scale is supported by literature showing that electro-thermal and ageing dynamics identified at the cell level can be extended to packs with proper calibration. Experimentally validated pack-level models indicate that voltage, temperature, and capacity degradation trends are preserved with errors below approximately 5\% compared to measured pack data, and cell-to-pack upscaling approaches have been shown to accurately predict pack-level lifetime under realistic operating conditions \cite{attia2020closed, calearo2022methodology, che2022lifetime, hosen2022strategic}. 

For a comprehensive overview, Table \ref{tab:parameters} presents the approximated parameters with scaling. The coefficients of the battery's cyclic capacity fading model, $k^{\rm{ec}}_1$, $k^{\rm{ec}}_2$, $k^{\rm{ec}}_3$, and $k^{\rm{ec}}_4$, were determined by fitting equation \eqref{eq:cycdeg} to the Stanford degradation dataset \cite{attia2020closed} using MATLAB's surface fitting tool, achieving a root mean squared percentage error of 3.7\%. Piece-wise linear relaxation was obtained by dividing SOC into increments of 2–20\% to determine coefficients for the degradation cost function in equation \eqref{eq:cycdeglinear} and the state-space formulation in equation \eqref{eq:statespaces1}. Table \ref{tab:rmsePWLM} reports normalized RMSE for the piece-wise state-space model and R$^2$ scores for the degradation cost function. Considering the increase in binary variables with finer increments, a 10\% SOC interval was selected, with the corresponding linearized coefficients given in Table \ref{tab:degrlinearapprox}.

\color{black}

\begin{table}[]
\footnotesize
\centering
\caption{Battery Parameters}
\label{tab:parameters}
{\renewcommand{\arraystretch}{1.0}
\begin{tabular}{|l|c|c|c|c|c|}
\hline
\textbf{Parameters} & $v_{\rm{OC}}$ & $Q$ & $R_0$       & $R_1$       & $R_1C_1$ \\ \hline
\textbf{Range}     & 3.2-4.2       & 2   & 0.004-0.014 & 0.002-0.008 & 60-250   \\ 
\textbf{Unit}  & V        & Ah & ohm       & ohm       & sec \\ 
\textbf{Scaling} & $b_sv_{\rm{OC}}$ & $b_p Q$          & $\frac{b_s}{b_p}R_0$ & $\frac{b_s}{b_p}R_1$ & $R_1C_1$      \\ \hline
\end{tabular}}
\end{table}

\begin{table}[]
\footnotesize
\centering
\caption{NRMSE and R2 for Different SOC Interval}
{\renewcommand{\arraystretch}{1.0}
\begin{tabular}{|l|c|c|c|c|c|}
\hline
\textbf{SOC increment} \textbf{(\%)} & \textbf{2} & \textbf{5} & \textbf{10} & \textbf{15} & \textbf{20} \\ \hline
NRMSE (\%) & 0.03 & 0.48 & 3.84 & 14.58 & 42.04 \\ 
R$^2$ & 0.987 & 0.986 & 0.984 & 0.978 & 0.971 \\ \hline
\end{tabular} }\label{tab:rmsePWLM}
\end{table}

\begin{table}[h]
\footnotesize
\centering
\caption{Range of coefficients for linear approximation of degradation function}
\label{tab:degrlinearapprox}
{\renewcommand{\arraystretch}{1.0}
\begin{tabular}{|c|c|c|c|c|}
\hline
$\mu$ & $i^{\rm{ec}}$      & $c^{\rm{ec}}_{0,l,\mu}$            & $c^{\rm{ec}}_{1,l,\mu}$             & $c^{\rm{ec}}_{2,l,\mu}$          \\ \hline
${\rm{CM_1}}$                                                    & [20,60]   & [-6e-4,-0.8e-4]  & [2.27e-6,9.15e-6]   & [1e-4,5e-4]     \\
${\rm{CM_2}}$                                                       & [80,200]  & [-5e-3,-6.6e-4]   & [6.34e-6,2.37e-5]    & [1.1e-3,4.3e-3] \\ 
${\rm{DM}}$                                                         & [-80,-20] & [-9e-4,-1.2e-4] & [-1.06e-5,-2.7e-6] & [2e-4,8e-4]     \\ 
${\rm{OFF}}$                                                        & 0       & 0                & 0                  & 0             \\ \hline
\end{tabular}}
\end{table}

\begin{table}[]
\footnotesize
\centering
\caption{Fleet description}
\label{tab:fleetdescription}
{\renewcommand{\arraystretch}{1.0}
\begin{tabular}{|c|c|c|c|}
\hline
  \textbf{\begin{tabular}[c]{@{}c@{}}Capacity \\ (kWh)\end{tabular}} &
  \textbf{\begin{tabular}[c]{@{}c@{}}Capacity \\ (Ah)\end{tabular}} &

  \textbf{\begin{tabular}[c]{@{}c@{}}Initial SOC  (\%) \end{tabular}} &
  \textbf{\begin{tabular}[c]{@{}c@{}}Batt cost \\ (\$/kWh) \end{tabular}} \\ \hline
  34.1 &
  100 &
  30-53& 300\\ \hline
\end{tabular}}
\end{table}

\subsection{Experimental Setup} \label{sec:expanalysis}
We have utilized a dataset containing 300 EVs (i.e., the number of vehicles that are stationed or en-route to the CS over a time horizon of 24~hr ) with varying vehicle-specific parameters as illustrated in Table \ref{tab:fleetdescription}. The dataset also contains the arrival time of the EVs at the charging station and initial SOC. The number of vehicles that are present simultaneously at the charging station may vary depending on the arrival time and customer requests. The case study focuses on price-based charging coordination, real-time operation scheduling, and the cost-effectiveness and robustness performance of the algorithm over the planning horizon. We compare the following scheduling algorithms: 

\begin{enumerate}
    \item \textbf{Business as usual (BAU):} BAU is a heuristic approach based on the current industry practice for fleet operators. The policy applies the maximum input current at each time slot during the charging period of each EV upon connection. 
   \item \textcolor{black}{\textbf{Laxity-based charging (LC):} The laxity-based policy \cite{nakahira2017smoothed} prioritizes vehicles according to their charging slack (time remaining until departure relative to the required charging time), giving higher priority to EVs with tighter deadlines to ensure timely SOC fulfillment.}
    \item \textbf{C-\icms:} This approach employs the Gurobi solver \cite{gurobi} for the centralized multi-objective multi-time step stochastic optimization problem in eq.~\eqref{eq:problemMPC} to determine the charging sequences at each time slot for all EVs.

    \item \textbf{D-\icms:} The hierarchical distributed problems in eq. \eqref{eq:problemMPC2} and \eqref{eq:problemMPC3} are solved using Gurobi \cite{gurobi} to determine the charging sequences at each time slot for individual EV.

\end{enumerate}

\begin{figure}[h]
\centering
\includegraphics[width=0.48\textwidth]{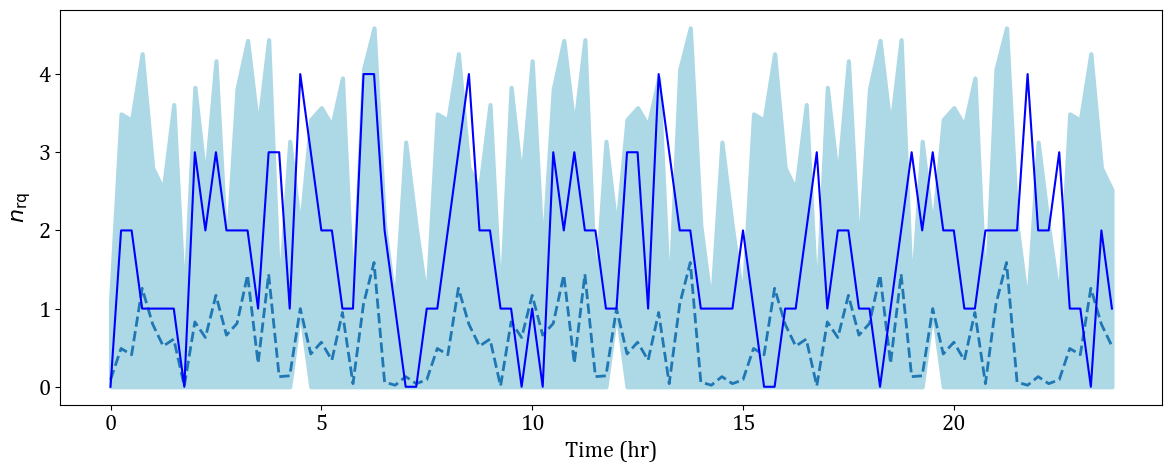}
\caption{Historical Demand, with solid and dotted lines, representing the actual requests and means of the request scenario, respectively and the spread of the plot, representing the maximum and minimum.}
\label{fig:demandscenario}
\end{figure}

To assess the energy efficiency and cost-effectiveness of the proposed \icms~framework, uniform Singapore energy prices (half-hourly energy price in the Singapore wholesale electricity market) \cite{emc2023} over a 24-hour horizon and an actual demand request from the historical demand at a charging station have been considered \textcolor{black}{along with the historical demand scenarios as shown in Fig.~\ref{fig:demandscenario}}. It is to be noted that the penalty term introduced to ease the energy requirement for ride allocation enables the algorithms to meet customer demand, even if the desired energy level has not been reached, by incurring an extra penalty cost ($\lambda_{\rm{pen}}$) of \$1/kWh for each kWh of energy unmet from the desired level. The switching costs among different charging modes are considered to be \$0.1 for \icms. The scaling factor $\lambda_{\rm{deg}}$ is found to be of the order of 40. The experiments have been performed in a Python environment on a notebook equipped with the Intel core-i7 processor and 16GB RAM.

\subsection{Comprehending Energy Consumptions and Charging Costs}
\begin{figure}[h]
    \centering
     \centering
         \includegraphics[width=0.49\textwidth]{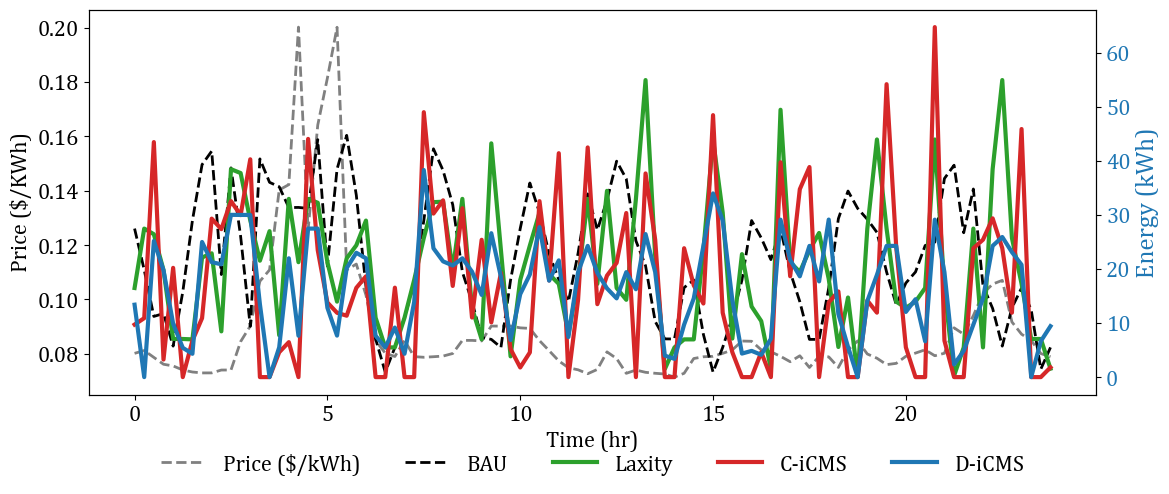}
     
    \caption{Price vs. energy consumption profile with $E_{\rm{all}}^{\rm{ec}}, \forall {\rm{ec}} \in \mathcal{EC}$ at 70\% of the maximum capacity with deterministic ride requests}
    \label{fig:costoptimality}
\end{figure}

\color{black}
To understand the cost-effectiveness, we assumed that the demand requests have been accurately predicted i.e., eq. \eqref{eq:problemMPC} and \eqref{eq:problemMPC2} has been treated as a deterministic optimization problem. Fig.~\ref{fig:demandscenario} showcases the demand profile scenarios and actual demand profile. The weight coefficients $w_1$ and $w_2$ from eq. \eqref{eq:problemMPC} and \eqref{eq:problemMPC3} are set as 0.9 and 0.1, respectively and ${E}_{\rm{all}}^{\rm{ec}}$ from eq. \eqref{eq:ready1} is set at 70\% of the maximum capacity. Fig. \ref{fig:costoptimality} demonstrate the comparative scheduling outcomes between the three strategies as described in section \ref{sec:expanalysis}. 

\color{black}

\color{black}

\begin{table}[h!]
\centering
\caption{Energy and cost efficiency analysis}
\label{tab:cost_energy_eff}
\begin{tabular}{|l|c|c|c|}
\hline
\textbf{Method} & \textbf{PAR} & \textbf{Elec. cost (\$)} &
 \textbf{Deg. cost (\$)} \\
\hline
BAU &  \textbf{2.04} &  193.95 & 21.04\\
Laxity &  2.68 & 176.54 & 19.63\\
C-\icms  &  3.89 & 137.45 & 7.90 \\
D-\icms  & 2.33 & 138.38 & 5.2\\
\hline
\end{tabular}
\end{table}

\color{black}

\color{black}
The performance comparison in Table~\ref{tab:cost_energy_eff} reveals clear trade-offs between peak demand characteristics and economic efficiency across different charging strategies. The BAU approach achieves the lowest PAR of 2.04, reflecting the smoothest load profile, but this comes with higher electricity and degradation costs, indicating suboptimal economic performance. The LC method lowers electricity cost to \$176.54 and slightly reduces degradation cost to \$19.63, yet exhibits a higher PAR of 2.68, showing that cost improvements come at the expense of increased load concentration.

Among the proposed strategies, the C-\icms~framework achieves the lowest electricity cost and relatively low degradation cost, demonstrating superior cost efficiency. However, this comes with a substantial increase in PAR, as the centralized algorithm aggressively exploits low-price periods (Fig.~\ref{fig:costoptimality}). In contrast, the D-\icms~approach strikes a balance, attaining near-optimal electricity cost, the lowest degradation cost, and a moderate PAR of 2.33.

These performance differences stem from the underlying charging philosophies. LC schedules charging based solely on SOC requirements and departure deadlines, without considering electricity prices, often consuming more energy during high-price periods. Both C-\icms~and D-\icms~incorporate price-informed charging via an MPC structure, which allows them to exploit temporal price variations and optimize energy consumption, resulting in lower electricity costs and occasional temporal spikes during low-price “valleys”.

The distinction between centralized and decentralized implementations further explains observed behaviors. C-\icms~leverages global system information to minimize cost but concentrates load, leading to a higher PAR. D-\icms~uses local information, slightly increasing electricity cost but achieving better load distribution and minimal degradation. While C-\icms~optimizes economic performance, D-\icms~offers a more balanced solution, and LC remains advantageous when peak reduction and grid stability are prioritized.
\color{black}

\subsection{Assessment of Stochastic Optimization Solution}

\begin{figure}[htbp]
    \centering
     \begin{subfigure}[Charging cost]{
     \centering
         \includegraphics[width=0.28\linewidth]{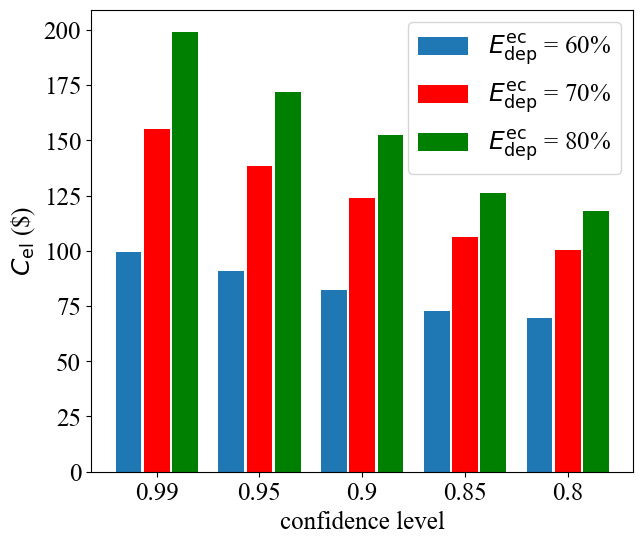}
         % \caption{Charging cost}
         \label{fig:alpha_e_cost}}
     \end{subfigure}  %
     \begin{subfigure}[Degradation cost]{
     \centering
         \includegraphics[width=0.28\linewidth]{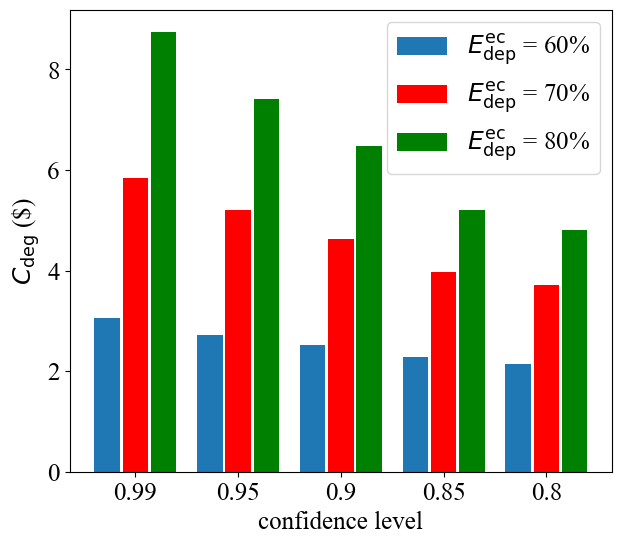}
         % \caption{Penalty violation}
         \label{fig:alpha_b_cost}}
     \end{subfigure}
      \begin{subfigure}[Penalty violation]{
     \centering
         \includegraphics[width=0.29\linewidth]{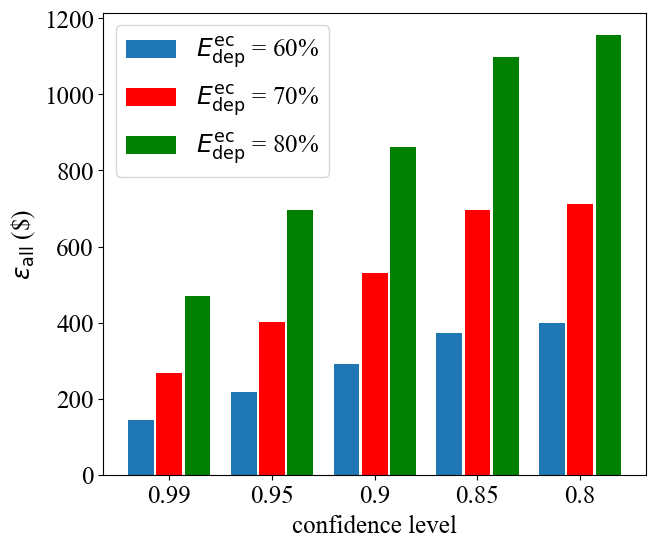}
         % \caption{Penalty violation}
         \label{fig:alpha_vio_cost}}
     \end{subfigure}

     \begin{subfigure}[deviation in charging cost]{
     \centering
         \includegraphics[width=0.28\linewidth]{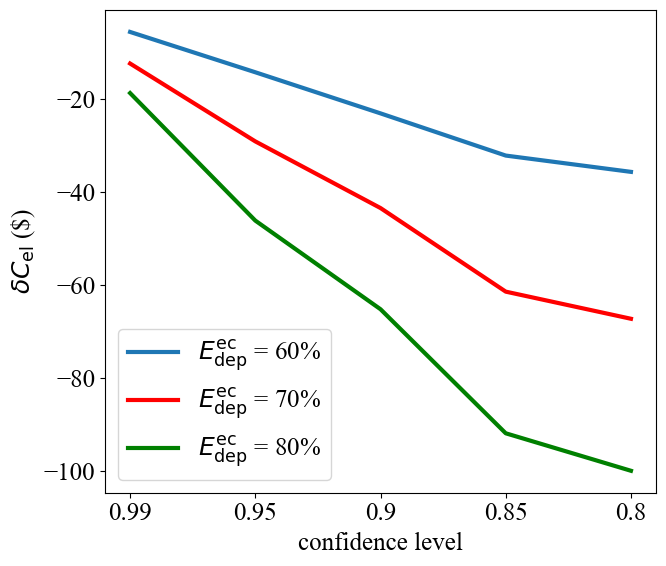}
         % \caption{deviation in charging cost}
         \label{fig:alpha_e_error}}
     \end{subfigure} %
     \begin{subfigure}[Deviation in degradation cost]{
     \centering
         \includegraphics[width=0.28\linewidth]{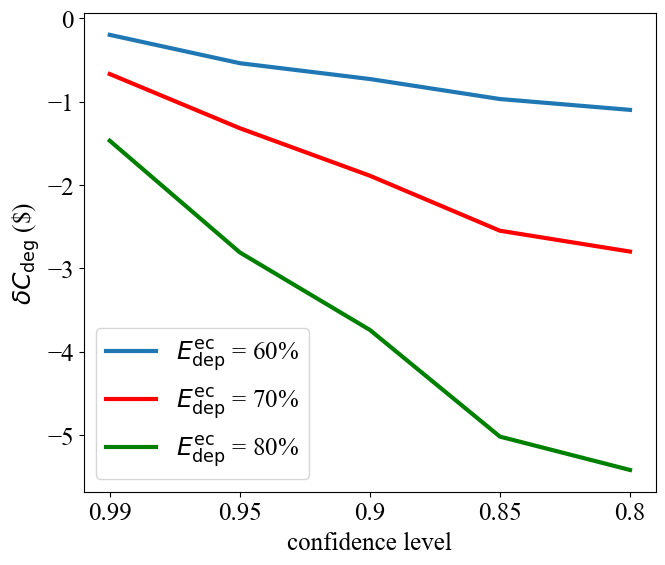}
         % \caption{Deviation in penalty}
         \label{fig:alpha_b_error}}
     \end{subfigure}
     \begin{subfigure}[Deviation in penalty]{
     \centering
         \includegraphics[width=0.28\linewidth]{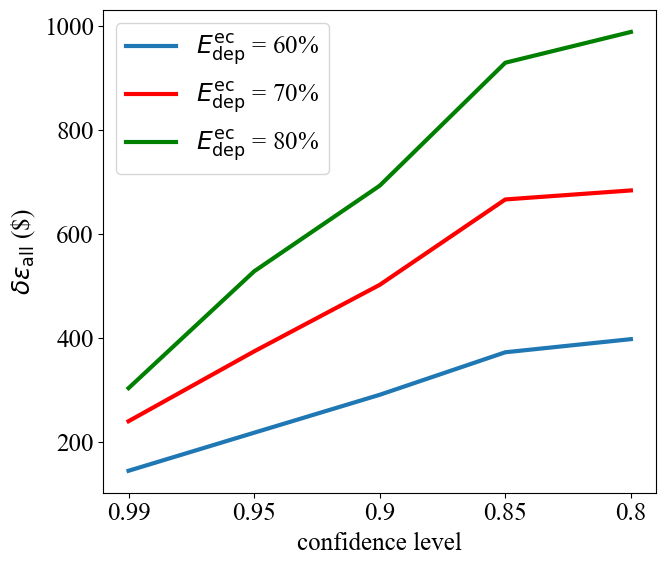}
         % \caption{Deviation in penalty}
         \label{fig:alpha_vio_error}}
     \end{subfigure}
    \caption{Stochastic optimization performance analysis for various confidence levels}
    \label{fig:confidencelevel}
\end{figure}

The stochastic objective function in eq. \eqref{eq:problemMPC} is evaluated using the proposed distributed hierarchical optimization algorithm. The solution accounts for the actual demand realization of the current time instant and \textcolor{black}{incorporates chance constraint formulation for the future time instants of the prediction horizon}. 
\color{black}
To assess the effectiveness of the proposed approach, the performance of stochastic optimization is examined in multiple charging modes and varying confidence levels, with $E^{\rm{ec}}_{\rm{all}}, \forall {\rm{ec}} \in \mathcal{EC}$ evaluated in the range of 60 to 80\%. The analysis is structured into two complementary sets of results. The first set, presented in Figs. \ref{fig:alpha_e_cost}–\ref{fig:alpha_vio_cost}, examines the influence of the confidence level on the optimized cost components. As the confidence level increases, constraint satisfaction is enforced over a broader set of demand scenarios, leading to more conservative decisions that are, in turn, more robust and less sensitive to demand uncertainty. Conversely, lower confidence levels permit the optimizer to exploit favorable realizations, which in turn reduce expected costs but simultaneously increase exposure to a greater number of unmet demands.

The second set of results, reported in Figs. \ref{fig:alpha_e_error}–\ref{fig:alpha_vio_error}, captures the cost difference between the optimized chance-constrained solution and the recourse action cost — defined here as the cost incurred under actual demand materialization, encompassing charging, battery degradation, and penalty cost components across the planning horizon. This difference reflects the implicit cost of uncertainty not fully anticipated by the chance-constrained formulation at a given confidence level, and serves as a key indicator of solution robustness. Higher confidence levels consistently yield smaller deviations between the chance-constrained and recourse action costs, confirming that more conservative decisions better anticipate actual demand outcomes and reduce the need for costly corrective adjustments. Nevertheless, the marginal reduction in this cost difference diminishes as the confidence level approaches unity, reflecting the characteristic trade-off inherent to chance-constrained optimization, whereby additional conservatism ceases to yield proportional gains in robustness against demand uncertainty. It is to be noted that higher confidence levels necessitate stricter service guarantees in the face of demand uncertainty, requiring greater fleet availability or charging flexibility. In contrast, lower confidence levels reduce operational burden at the expense of an increased likelihood of unmet ride requests.

\color{black}
\subsection{Multi-Objective Framework Assessment}
\begin{figure}[]
\centering
\includegraphics[width=0.33\textwidth]{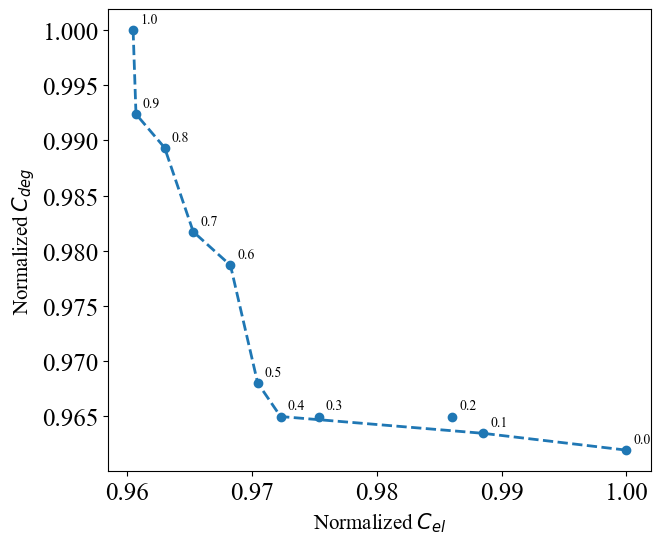}
\caption{Pareto front with different weight vectors for charging and degradation cost. The weight $w_1$ is indicated.}
\label{fig:paretofront}
\end{figure}

Fig. \ref{fig:paretofront} illustrates the impact of favoring one objective over another through the allocation of greater weights. To understand the effect of optimizing one objective more over others, we have varied the weight $w_1$ from 1 to 0 to capture the optimized objectives. As evident from Fig. \ref{fig:paretofront}, applying a higher weight to the charging cost yields the most pronounced reduction in charging expenses compared to the other alternative weight combinations. The battery degradation cost has been seen to be improved with a weight of 0.1 compared to the lower weight of 0.9. Although the absolute value of the degradation cost related to battery value preservation as reported in Fig. \ref{fig:paretofront} might not be considerably substantial, its long-term implications prove beneficial in addressing battery state of health, resale value, and collective driving range of the fleet. Thus, the incorporation of a comprehensive degradation model followed by its piece-wise linearization for improved computational complexity into the framework captures the degradation of the EV battery with higher accuracy.

\subsection{Understanding the Effect of Hyperparameters}

\begin{figure}[]
\centering
\includegraphics[width=0.4\textwidth]{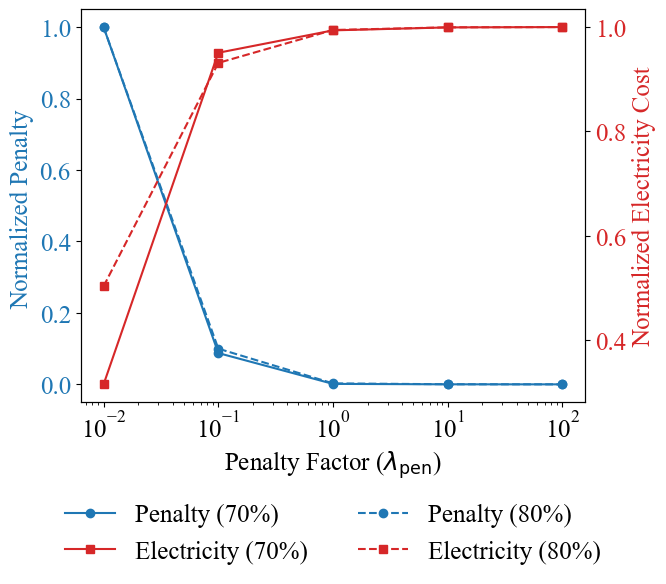}
\caption{Normalized penalty cost and electricity cost analysis at different penalty factor and $E_{\rm{all}}^{\rm{ec}}$ level}
\label{fig:penalty}
\end{figure}

The model formulation involves two key hyper-parameters: (i) penalty cost and (ii) switching cost. The penalty cost significantly impacts the energy state of electric vehicles (EVs) at departure. Fig.~\ref{fig:penalty} illustrates the relationship between charging costs and penalty violation costs across various penalty factors. At the lowest penalty factor of 0.001, the charging cost is minimized, but this results in more penalty violations, causing vehicles to depart with insufficient SOE. Conversely, a higher penalty cost reduces penalty violations but leads to increased energy costs.

\begin{table}[]
\footnotesize
\centering
\caption{Understanding the effect of switching parameters}
\label{tab:switch}
{\renewcommand{\arraystretch}{1.0}
\begin{tabular}{|c|c|c|c|c|}
\hline
\textbf{Variations} &
  \textbf{PAR} &
  \textbf{Elec. cost(\$)} &
  \textbf{Deg. cost(\$)} &
  \textbf{Penalty(\$)} \\ \hline
D-\icms    & 5.05 & 164.58 & 10.73 & 1.55 \\
D-\icms-V1 & 2.6 &166.5 & 6.88 & 14 \\
D-\icms-V2 & 5.15 & 173.6 & 14.66 & 1.55\\ \hline
\end{tabular}}
\end{table}
\begin{figure}[h]
\centering
\includegraphics[width=0.5\textwidth]{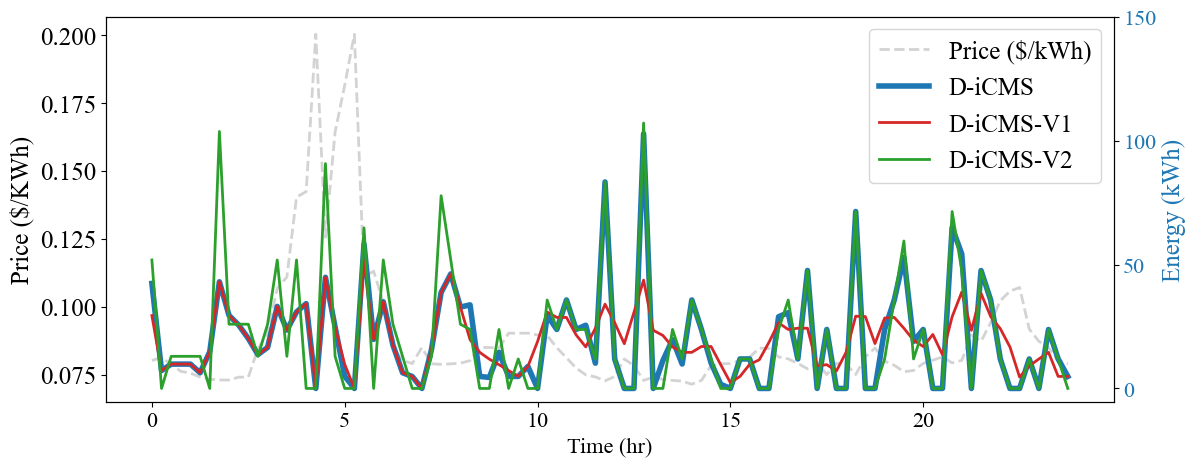}
\caption{Consumption profile for variations in switching parameters}
\label{fig:switch}
\end{figure}

In addition, we have experimented with the following combinations of switching costs to understand their impact on the overall operational, degradation cost, and penalty violation cost:

\begin{enumerate}
    \item D-\icms-V1: by setting $c^{\rm{ec}}_{\rm{OFF, CM_2}}$, $c^{\rm{ec}}_{\rm{CM_2, OFF}}$, $ c^{\rm{ec}}_{\rm{CM_2, CM_1}}$, $ c^{\rm{ec}}_{\rm{CM_1, CM_2}} $ at higher values of $\$10$;
    \item D-\icms-V2: by setting $c^{\rm{ec}}_{\rm{OFF, CM_1}}$, $c^{\rm{ec}}_{\rm{CM_1, OFF}}$, $ c^{\rm{ec}}_{\rm{CM_2, CM_1}}$, $ c^{\rm{ec}}_{\rm{CM_1, CM_2}} $ at higher values of $\$10$.
\end{enumerate}
\color{black}
Table~\ref{tab:switch} summarizes the electricity costs, battery degradation costs, penalty terms, and PAR at the 70\% $E_{\rm{all}}^{\rm{ec}}$ level for different switching configurations. To clearly highlight the impact of switching parameters, the maximum power draw constraint is relaxed in this analysis, allowing the effects to be observed more prominently without loss of generality.

The D-\icms-V1 variant, characterized by switching costs that favor slower charging, achieves a substantially lower PAR compared to the other configurations. This reduction in charging intensity also leads to significantly lower battery degradation costs. However, this conservative charging behavior results in a notable increase in penalty costs, indicating more frequent constraint violations. In contrast, D-\icms-V2, which relies exclusively on fast charging, exhibits the highest degradation cost and a slightly higher PAR, reflecting more aggressive energy usage. Nevertheless, it maintains low penalty costs, as the higher charging rates enable better compliance with operational constraints. The baseline D-\icms~strikes a balance between these two extremes, achieving moderate electricity cost, degradation cost, and low penalty cost, while maintaining a PAR of 5.05. It is to be noted that the results signify that the unified model can be tuned via switching cost parameters to prioritize either grid smoothness and battery health (D-\icms-V1) or operational feasibility and constraint satisfaction (D-\icms-V2), with D-\icms\ providing a balanced trade-off among the competing objectives.

\color{black}

\subsection{Run-time Analysis} \label{sec:complexity}

\color{black}
Table \ref{tab:runtime} presents the runtime of C-\icms~and D-\icms~approaches under different sampling intervals and prediction horizons for a 24-hour planning horizon. Given a price signal with a 30-minute sampling interval, a first-order hold method is employed to model intermediate points when using shorter sampling intervals. As expected, the runtime of the C-\icms~approach increases significantly with shorter sampling intervals and longer prediction horizons, primarily due to the growth in the number of binary variables and the number of linear programming subproblems generated through cutting planes, branching, and bounding nodes. This highlights the scalability challenges of centralized implementation for real-time operation. In contrast, the D-\icms~approach decomposes the problem into a two-stage process. The higher-level stage solves a simplified assignment problem, while the lower-level stage distributes the detailed computation among individual EV agents, reducing the number of decision variables per subproblem and enabling parallel computation. The tabulated runtime corresponds to the combination of the higher-level solution and the maximum runtime among all lower-level EV agents, illustrating the computational efficiency of the distributed framework.

\begin{table}[]
\centering
\caption{Runtime (in sec) w.r.t. prediction horizon and sampling interval}
\footnotesize
\label{tab:runtime}
{\renewcommand{\arraystretch}{1.2}
\begin{tabular}{|l|cccc|}
\hline
\multicolumn{1}{|c|}{\multirow{2}{*}{\textbf{Algorithm}}} &
  \multicolumn{4}{c|}{\textbf{Interval(min)}} \\ \cline{2-5} 
   &
  \multicolumn{1}{c|}{\textbf{10}} &
  \multicolumn{1}{c|}{\textbf{15}} &
  \multicolumn{1}{c|}{\textbf{20}} &
  \multicolumn{1}{c|}{\textbf{30}}  \\ \hline
\multirow{1}{*}{\textbf{C-\icms}}  &
  \multicolumn{1}{c|}{190.25} &
  \multicolumn{1}{c|}{20.38} &
  \multicolumn{1}{c|}{27.8} &
  3.2 \\
   \hline
\multirow{1}{*}{\textbf{D-\icms}}  &
  \multicolumn{1}{l|}{3.8} &
  \multicolumn{1}{l|}{1.35} &
  \multicolumn{1}{l|}{3.8} &
  \multicolumn{1}{l|}{1.3} 
 \\ \hline
\end{tabular}}
\end{table}

\begin{table}[]
\footnotesize
\centering
\color{black}
\caption{\color{black}Runtime and degradation cost deviation analysis for D-\icms~under different SOC discretizations}
\label{tab:runtime_soc}
{\renewcommand{\arraystretch}{1.2}
\begin{tabular}{|c|c|c|c|c|}
\hline
\textbf{Metric} & \textbf{2\%} & \textbf{5\%} & \textbf{10\%} & \textbf{20\%} \\ \hline
Runtime (s)      & 223.7 & 20.2 & 1.35 & 0.41 \\ \hline
Deg. Dev. (\%)   & 0.04  & 0.19 & 0.52 & 1.93 \\ \hline
\end{tabular}}
\end{table}

Table \ref{tab:runtime_soc} further explores the impact of SOC discretizations on runtime and battery degradation cost deviation in the D-\icms~framework. Finer SOC increments lead to a larger number of binary variables, which substantially increases runtime, whereas coarser increments reduce computational effort but slightly increase the deviation in degradation cost, calculated as the difference between the original nonlinear degradation model in \eqref{eq:cycdeg} and the piecewise linear approximation from \eqref{eq:cycdeglinear} used in the simulation. For instance, small SOC increments yield the most accurate representation of battery aging but result in the highest runtime, which may limit practical applicability in real-time or large-scale deployments. Conversely, moderate to coarse SOC increments dramatically reduce runtime while maintaining degradation cost deviation within acceptable bounds, highlighting a key trade-off between computational efficiency and modeling accuracy.  The centralized C-\icms~method achieves an optimality gap of about 0.1\%, while the distributed D-\icms~method reaches an even smaller gap of roughly 0.005\%, demonstrating that both approaches yield near-optimal solutions.

\section{Conclusion} \label{sec:conclusion}
\color{black}
This study proposes an integrated charge-scheduling and service-allocation framework to improve service quality and reduce operational costs for EV ride-hailing customers. We formulated a stochastic Mixed-Integer Linear Programming Model Predictive Control (MILP-MPC) framework, supported by a hierarchical distributed algorithm to ensure computational scalability. The analysis demonstrates that our D-\icms~framework achieves a favorable balance between real-time implementability and solution fidelity. While 10\% SOC discretizations offer a robust starting point, careful tuning remains essential to balance runtime performance with battery degradation accuracy as fleet sizes grow.

Finally, the framework provides a practical decision-support tool for fleet management. By adjusting the weights of the cost function and cost coefficients, fleet managers can achieve the following objectives:
\begin{itemize}
    \item \textbf{balance competing objectives:} Real-time tuning between electricity costs, service reliability, and battery longevity allows for operational adaptability to changing market conditions;
    \item \textbf{enable grid-Responsive dispatching:} The model facilitates grid-conscious charging, helping to mitigate peak-load impacts while taking advantage of potential energy price incentives;
    \item \textbf{scale sustainably:} The decentralized computation structure allows managers to scale fleet operations independently of the computational bottlenecks common in centralized systems.
\end{itemize}
\color{black}

\section*{Acknowledgment}
This study is supported under the RIE2020 Industry Alignment Fund – Industry Collaboration Projects (IAF-ICP) Funding Initiative, as well as cash and in-kind contributions from the industry partner(s). \textcolor{black}{This research is also part of the programme DesCartes and is supported by the National Research Foundation, Prime Minister’s Office, Singapore, under its Campus for Research Excellence and Technological Enterprise (CREATE) programme.}

\bibliographystyle{IEEEtran}
\bibliography{ref}

\begin{IEEEbiography}[{\includegraphics[width=1in,height=1.25in,clip,keepaspectratio]{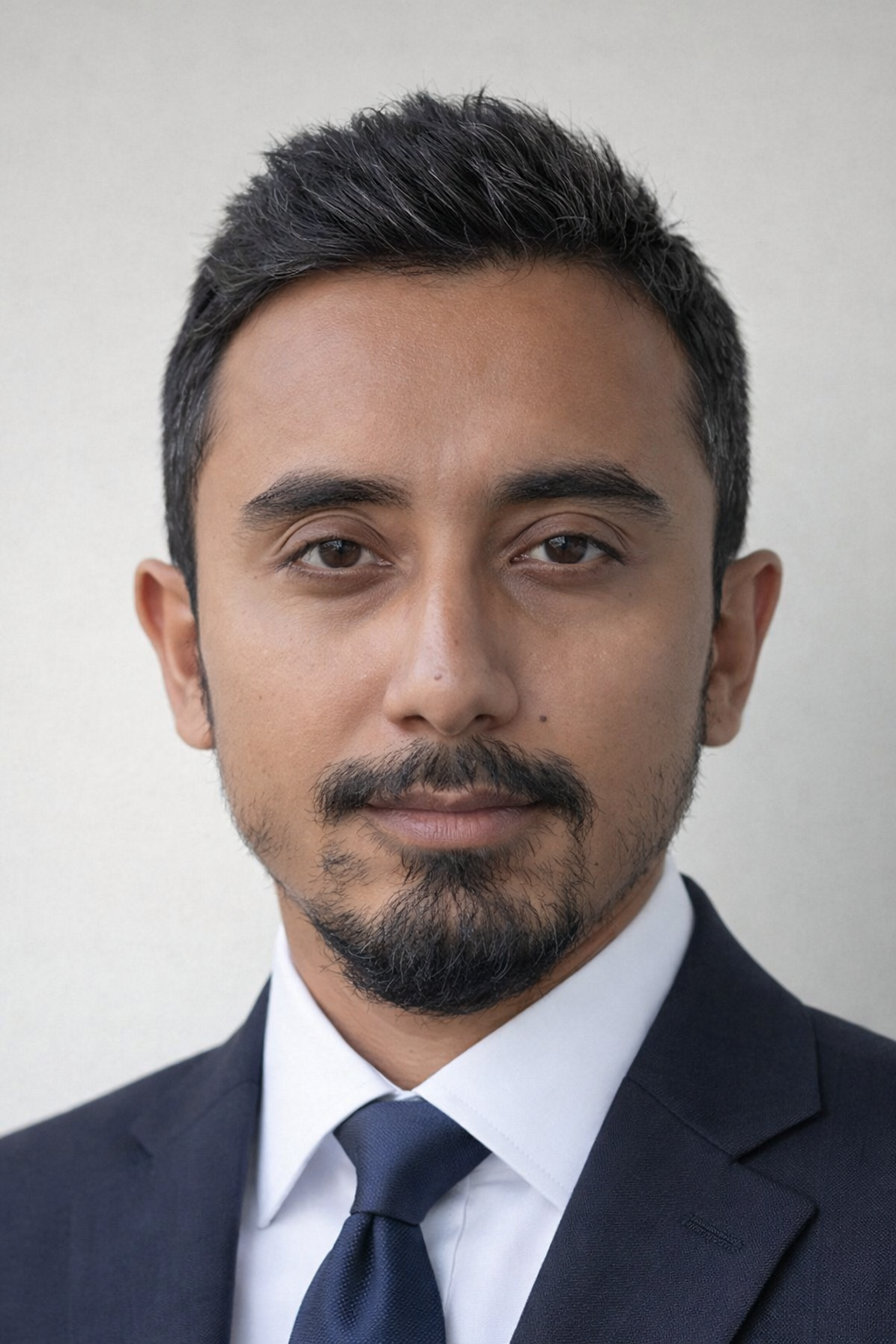}}]%
{Mainak Dan}
received the B.E. and M.E. degrees in Electronics and Communication Engineering from Jadavpur University, Kolkata, India, and the Ph.D. degree from Nanyang Technological University, Singapore. He is currently a Research Engineer at G2Elab, Grenoble INP, France, where he focuses on the development of open-source optimization and simulation tools for power and multi-energy systems, specifically within the framework of EU-funded projects. His expertise spans the stochastic optimization of energy systems, the strategic integration of storage technologies, electric vehicle fleet charging optimization, and the management of distributed energy resources. His research interests are centered on the application of advanced mathematical optimization methods to solve complex, real-world energy challenges and facilitate sustainable energy transition.
\end{IEEEbiography}

\begin{IEEEbiography}[{\includegraphics[width=1in,height=1.25in,clip,keepaspectratio]{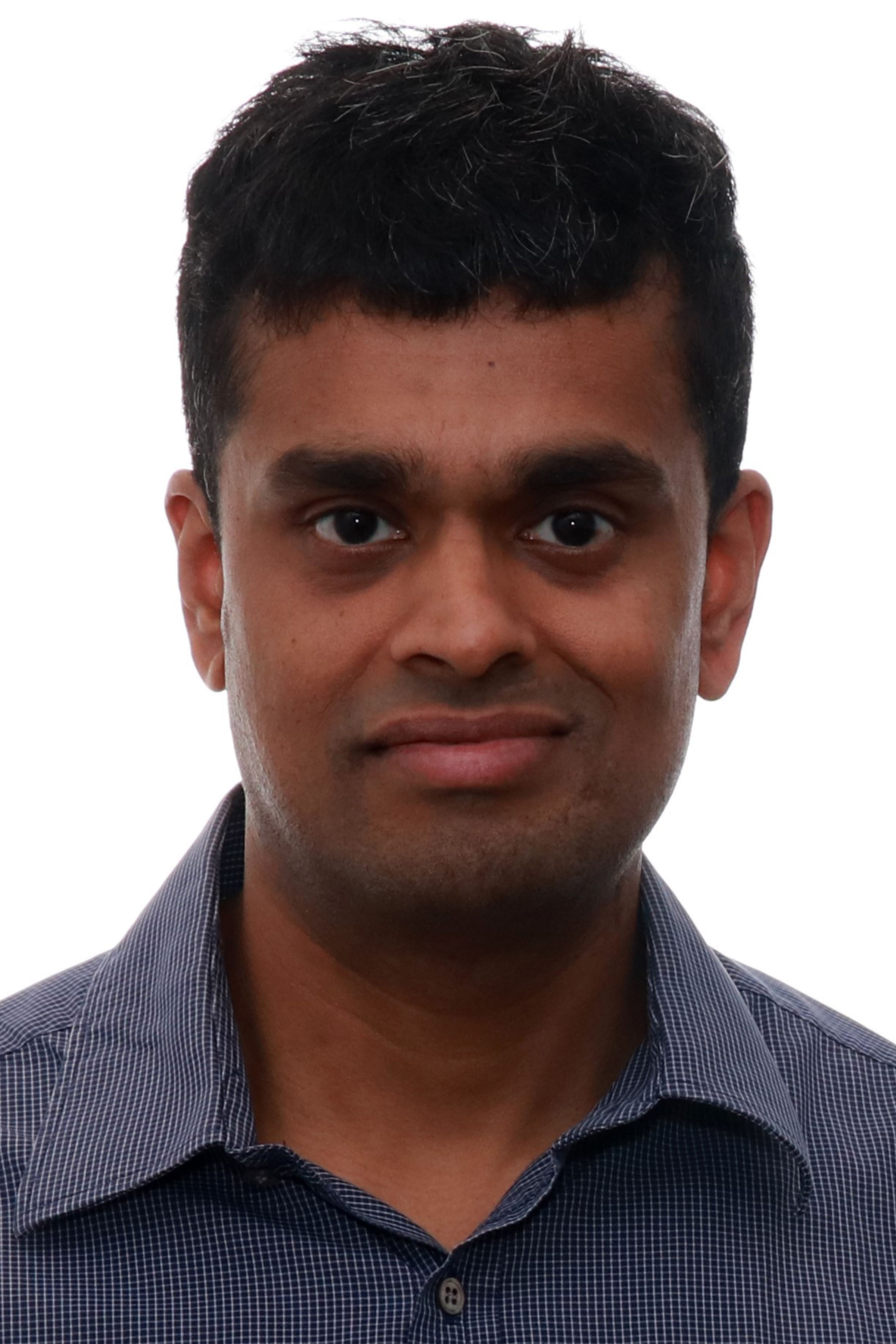}}]%
{Arvind Easwaran} is an Associate Professor and Assistant Dean (Academic) in the College of Computing and Data Science at Nanyang Technological University (NTU), Singapore. He is also the co-Director for the CNRS@CREATE program DesCartes on AI-based robust decision-making for urban critical systems. He received a PhD degree in Computer and Information Science from the University of Pennsylvania, USA, in 2008. Prior to joining NTU in 2013, he has been an Invited Research Scientist at the Polytechnic Institute of Porto, Portugal, in 2009-2010, and a Research \& Development Scientist at Honeywell Aerospace, USA, in 2010-2012. In NTU, his research focuses on the design and analysis of real-time and cyber-physical computing systems, including their application in domains such as automotive and urban energy systems. He is a Senior Member of IEEE.
\end{IEEEbiography}

\end{document}